\documentclass[fleqn,usenatbib]{mnras}

\usepackage{newtxtext,newtxmath}
 
\usepackage[utf8]{inputenc}

\usepackage[T1]{fontenc}
\usepackage{ae,aecompl}
\usepackage{lineno}
\usepackage{graphicx}	% Including figure files
\usepackage{amsmath}	% Advanced maths commands
\usepackage{amssymb}	% Extra maths symbols
\usepackage{soul} % Editing package
\usepackage{tablefootnote}
\usepackage{subfig}
\usepackage{lipsum}

\RequirePackage{color}

\DeclareUnicodeCharacter{0301}{\'{}}  % Acute accent
\title[LSBds around NGC\,3115]{Low surface brightness dwarf galaxies and their globular cluster populations around the low-density environment of our closest S0 NGC\,3115}

\author[Canossa-Gosteinski, Chies-Santos, Furlanetto et al.]{Marco A. Canossa-Gosteinski$^{1}$\thanks{canossa.marco@gmail.com}, Ana L. Chies-Santos$^{1}$\thanks{ana.chies@ufrgs.br}, Cristina Furlanetto$^{1}$\thanks{cristina.furlanetto@ufrgs.br}, 
\newauthor 
Charles J. Bonatto$^{1}$,
Rodrigo Flores-Freitas$^{1}$, William Schoenell$^{1,2}$, 
Michael A. Beasley$^{3,4,5}$,\newauthor
Roderik Overzier$^{6,7}$, 
Basilio X. Santiago$^{1}$,
Adriano Pieres$^{1,8}$,
Em\'ilio J. B. Zanatta$^{1,10}$,
\newauthor 
Karla A. Alamo-Martinez$^{1,9}$,
Eduardo Balbinot$^{11}$,
Anna B. A. Queiroz$^{4, 5}$,
Alan Alves-Brito$^{1}$
\\
$^{1}$Instituto de F\'isica, Universidade Federal do Rio Grande do Sul, Av. Bento Gonçalves 9500, Porto Alegre, RS, Brazil\\
$^{2}$GMTO Corporation 465 N. Halstead Street, Suite 250 Pasadena, CA 91107, USA \\
$^{3}$Centre for Astrophysics and Supercomputing, Swinburne University, John Street, Hawthorn VIC 3122, Australia\\
$^{4}$Instituto de Astrofísica de Canarias, Calle Vía Láctea, E-38206 La Laguna, Spain\\ 
$^{5}$Departamento de Astrofísica, Universidad de La Laguna, E-38206 La Laguna, Spain\\
$^{6}$Observatório Nacional, Rua General José Cristino, 77, São Cristóvão, 20921-400 Rio de Janeiro, RJ, Brazil\\
$^{7}$Leiden Observatory, Leiden University, Niels Bohrweg 2, 2333 CA, Leiden, The Netherlands\\
$^{8}$Laboratorio Interinstitucional de e-Astronomia - LIneA, Av. Pastor Martin Luther King Jr, 126 - Torre 3000/817, Rio de Janeiro, RJ - 20765-000, Brazil\\
$^{9}$Departamento de Astronoḿıa, Universidad de Guanajuato, Apartado Postal 144, 36000 Guanajuato, Guanajuato, Mexico\\
$^{10}$Universidade de São Paulo, IAG, Rua do Matão 1226, Cidade Universitária, São Paulo 05508-900, Brazil\\
$^{11}$Kapteyn Astronomical Institute, University of Groningen, Landleven 12, 9747 AD, Groningen, The Netherlands\\
}
\date{Accepted 2024 September 10. Received 2024 September 10; in original form 2023 November 17}

\pubyear{2024}

\begin{document}
\label{firstpage}
\pagerange{\pageref{firstpage}--\pageref{lastpage}}

\maketitle

%Abstract of the paper
\begin{abstract}
Understanding faint dwarf galaxies is fundamental to the development of a robust theory of galaxy formation on small scales. Since the discovery of a population of ultra diffuse galaxies (UDGs) rich in globular clusters (GCs) in Coma, an increasing number of studies on low surface brightness dwarf galaxies (LSBds) have been published in recent years.
The most massive LSBds have been observed predominantly in groups and clusters, with properties displaying dependence on the environment. In this work, we use deep DECam imaging to systematically identify LSBds and their GC populations around the low-density environment of NGC 3115. We carefully analyse the structure and morphology of 24 candidates, 18 of which are reported for the first time.  
Most candidates exhibit red colours suggesting a connection between their colour and distance to NGC 3115.  
We followed up with Gemini GMOS imaging 9 LSBds to properly identify their GC populations. We derive lower limits for the number of GCs associated with each galaxy. Our analysis reveals that they occur around of the same loci of Fornax LSB dwarf GC systems. 
The relationship between the number of GCs and total mass provides a tool in which, by counting the GCs in these galaxies, we estimate an upper limit for the total mass of these LSB dwarfs, obtaining the mean value of $\sim 3.3\times10^{10}$ M$_{\odot}$. Our results align with expectations for dwarf-sized galaxies, particularly regarding the distribution and specific frequency of their GC systems. 
 
\end{abstract}

\begin{keywords}
galaxies -- dwarf galaxies -- elliptical and lenticular --  star clusters: individual
\end{keywords}

%%%%%%%%%%%%%%%%%%%%%%%%%%%%%%%%%%%%%%%%%%%%%%%%%%

%%%%%%%%%%%%%%%%% BODY OF PAPER %%%%%%%%%%%%%%%%%%

\section{Introduction}

The search for small and faint satellite dwarf galaxy systems has gained significant importance over the last few years due to their relevance to cosmology, particularly the missing satellites problem coupled with advancements in instrumentation and observing techniques. This has led to a growing interest in studying these objects, as they provide valuable insights into constraining galaxy formation on smaller scales. The finding and characterization of faint dwarf galaxies offer crucial constraints for theories of galaxy formation and $\Lambda$ CDM on small scales \citep{bbk2017}. Substantial progress has been made in both observational \citep{simon19,prole19} and theoretical \citep[e.g.][]{wetzel16, buck19, Martin19} aspects. 

The missing satellite problem stems from a disconnection between the numerous dark matter halos predicted by cosmological simulations, each potentially capable of hosting a dwarf galaxy, and the relatively small fraction of these galaxies that are actually observed. This issue becomes particularly pronounced when we consider the Milky Way, where there exists a substantial mismatch between the expected number of dwarf galaxies according to theoretical predictions and what is observed \citep{Klypin99, Moore99, Nashimoto22, Muller23}. This persistent discrepancy remains unsolved. Exploring these elusive, faint satellite dwarf galaxies provides vital insights into the processes of galaxy formation, evolution, and the fundamental nature of dark matter itself. Resolving this problem has the potential to significantly advance our comprehension of how galaxies form across all scales, contributing a crucial piece to the intricate puzzle of cosmological structure formation.

Low surface brightness dwarf galaxies (LSBds) are present in a wide range of physical sizes and environments. These objects have stellar masses comparable to those found in dwarf galaxies [M$_{\star} \leq 10^9$ M$_{\odot}$ \citealt{Prole21b}], but the distinguishing feature is their more diffuse light distribution. Additionally, they display  $r$-band surface brightness fainter than 24 magnitudes per square arc-second \citep{Martin19, prole19, carlsten2020a, saifollahi22}.

Ultra-faint dwarfs are generally faint ($-2.2 < M_V < -7.4$), have physical sizes ranging from $10 - 170 $ pc \citep{bechtol15, simon19} and are better studied in the Local Group.
The ultra-faint satellite census and characterization of the Milky Way and Local Group \citep{mac12, simon19} through resolved stars studies have yielded valuable insights. However, expanding to larger distances at the Local Volume is challenging \citep[e.g.][]{chiboucas13, lee2017} and so far has been largely dependent on diffuse integrated light searches \citep{bennet17, danieli18, carlsten2020a}.

LSBd galaxies beyond the Local Group have been known for over 30 years \citep[e.g.][]{SB84,cb1987, dalcanton97, wittmann17}, initially in small numbers.
The observations of over 40 such galaxies in the Coma Cluster \citep{vD15} by the Dragonfly Telephoto array \citep{abraham14} spiked a hunt in the extragalactic community for these low surface brightness galaxies \citep{vdB16, yagi16}. These recently uncovered LSB dwarf galaxies have $R_e \gtrsim 1.5$\,kpc and $M_V \lesssim -15$ (stellar mass, $M_\star \gtrsim 10^7 M_\odot$) and are now commonly found in the literature as ultra diffuse galaxies (UDGs) \citep{koda15, mihos15, mancera19}. However, the continuum in properties from UDGs to smaller LSB dwarfs, coupled with their shared similar stellar populations \citep{Fensch19}, suggest that they are the same population only differing in size \citep{conselice18}.

UDGs found in high-density environments are typically red, have many GCs \citep{beasley16, Lim2018, prole19a} and old stellar populations ($>7$\,Gyrs) \citep{ruizlara18, pandya18, Mateu18}. When systematic searches are performed in the field, low surface brightness dwarf galaxies appear as blue, clumpy and star-forming \citep{bell17, zaritsky19, prole19, Martin19, Tanoglidis20}. Some UDGs in the field also display HI emission \citep{k20}, with a few of this objects being gas-rich. The wide range of physical properties implies that there are many possible ways and processes to explain the formation and evolution of these galaxies.

Most of the large LSB dwarf galaxies seem to be dark matter dominated \citep{penny09, saifollahi22} that formed in dark matter halos with high angular momentum \citep{amorisco16}.
There are a few channels for the formation of large LSB dwarf galaxies. 
One interpretation is that they are galaxies that have been transformed by external processes, such as ram-pressure and tides at cluster infall \citep{yozin15, rong17, carleton19, sales20, Watkins23}. Another explanation relies on internal factors and suggests they form through bursty star formation and episodic supernovae outflows
\citep{dicintio17, Jiang18, chan18}.

The study using cosmological simulations conducted by \citet{Martin19}, aimed at understanding the formation and evolution mechanisms of LSBds, highlighted that external factors, like ram-pressure stripping, play a significant role in driving the evolution of these systems. LSBds make a substantial contribution to the overall galaxy density, accounting for 47\% of the local number density. However, despite their high numbers, their impact on total stellar mass and luminosity is quite the opposite, comprising only 7\% and around 6\% respectively \citep{Martin19}. 

In another recent study on the formation and evolution of UDGs using cosmological simulations conducted by \citet{Benavides23}, it was found that the majority of UDG formation in the simulations occurs due to internal processes. Notably, this formation process is characterized by a significant contribution from high-spin dark matter haloes.

A key question that remains to help distinguish between models is whether the red UDGs have counterparts in host halos with lower masses and, if so, what are their abundances as a function of halo mass. \citet{vdB17} shows that compared to bright galaxies, UDGs are relatively more abundant in massive clusters \citep[e.g.][]{lee2020} than in groups. Their work shows that it is still unclear whether this difference is related to a higher destruction rate of UDGs in groups or if massive halos have a positive effect on UDG formation. To date, there have been only a few large LSBs reported in low-density environments \citep[e.g.][]{ft2016, md2016, roman2019} that are red. 
Yet to be understood is the significant difference between the number of field and cluster UDGs. Is it due to a mechanism that makes UDGs more likely to form or survive in high-density environments, or is it related to an observational bias? \citep{roman_trujillo2017}.

\citet{Prole21a} studied the quiescent fraction of isolated LSB dwarfs using data obtained from Hyper Suprime Camera (HSC) and Galaxy And Mass Assembly \citep[GAMA]{baldry2010} spectroscopy. They find that blue LSB dwarfs exist predominantly in low-density environments and that the red LSBG population is spatially correlated with local structures.
The bluer population tends to have a higher Sérsic index and more concentrated profiles. They
find around 26$\pm 5\%$ of isolated local LSB dwarfs belonging to the red population, indicating that high-density environments could play a dominant - but not exclusive - role in producing quiescent 
LSB dwarf galaxies.

Globular clusters (GCs) offer an interesting opportunity to further investigate processes that formed LSB galaxies. GCs are formed at early epochs \citep[e.g.][]{chies11, beasley20} and their properties are strongly connected to the assembly histories of their hosts. 
The physical mechanisms that shape the origin of globular clusters and dictate if they will be destroyed or survive \citep{Kruijssen15,Choksi19} and the following accretion episodes that give rise to current GC systems \citep{Bica06, Forbes10, Forbes11, Beasley2018} come hand-in-hand with the evolution of their host galaxies \citep{RodriguezGomez16, Davison20}. This is supported by the constant GC-to-halo mass relation, described in both observational \citep[e.g.][]{Spitler09,Hudson2014,harris2017} and numerical studies \citep[e.g.][]{El-Badry2019}. 

Thus, GCs are relevant for understanding the hierarchical assembly processes, because they are found around galaxies spanning a large range of masses, from dwarfs to giants \citep{Strader2005, beasley20}, and are discrete, bright beacons that help shed light on the evolution of their host galaxies. In addition to their high intrinsic brightness, another property makes them of key interest to galaxy evolution studies. Having mean ages older than $\sim$10 Gyrs \citep{Strader2005, chies11} GCs act as fossil tracers of galaxy evolution and its environment. Moreover, GCs serve as effective tracers of old stellar populations and provide a means to estimate the total mass of their host galaxies, thus establishing a crucial link for inferring the presence and quantity of dark matter halos.

To understand the formation and evolution of the few known low surface brightness galaxies found in low-density environments, we need a comprehensive study of their GC populations \citep{vanDokkum19, Lim2018}. The number of GCs correlates well with halo mass within a given galaxy system \citep{Blakeslee1997, harris2017, zaritsky22}. Moreover, halo masses themselves correlate well with the total mass of its GC system \citep{El-Badry2019, Hudson2014}.  Counting GCs may even offer ways of determining virial masses for (massive) LSB galaxies, and from the large numbers of GCs found in such systems, a quenching scenario that happened (at $z\sim3$) after the bulk of GC formation is favoured \citep{beasley16}. Several recent studies have since studied GC systems in LSB galaxies \citep[e.g.][]{Lim2018,amorisco18, prole19a, Muller2021, marleau21, saifollahi22}. 

To shed some light on how much low surface brightness galaxies properties are dependent on their environment we have obtained deep BLANCO/DECam (and Gemini/GMOS) imaging around the nearby galaxy NGC3115 (with a local surface density of $\Sigma \sim 20 \rm Mpc^{-2}$ for $M_B \lesssim -12$, \citealt{karachentsev2014}) and performed a systematic search for low surface brightness objects (and their GC systems) around our observed fields.
We note that our analysis is independent from that of \cite{carlsten21a} performed with DECaLS.

This paper is structured as follows. In Section \ref{sec:data} we describe the observations and data calibration procedures. In Section \ref{sec:visua_inspect} we describe our visual inspection procedure and the tool we built for this purpose. In Section \ref{sec:analysis} we describe our photometric and structural analysis and in Section \ref{sec:GCs} we analyse the systems with GC candidates. 
In Section \ref{sec:discussion} we discuss our results and summarize our findings in Section \ref{sec:summary}. 
All magnitudes we quote are in the AB  system.  Throughout this paper, we assume a distance modulus $(m-M) = 29.93 \pm 0.09$ mag for NGC 3115 \citep{Tonry01} corresponding to a distance of 9.7 $\pm$ 0.4 Mpc. We adopt the $\Lambda$CDM cosmological parameters with $H_0=73$ km s$^{-1}$ Mpc$^{-1}$, $\Omega_{\Lambda}=0.73$ and $\Omega_M=0.27$. \\

\section{Data}
\label{sec:data}
In this Section we describe the data used in this work: DECam and follow-up Gemini/GMOS imaging.

\subsection{Blanco/DECam imaging}
\label{sec:blanco}

We conducted observations in 2017 during four half nights (February 15–18) using the Dark Energy Camera (DECam) on the prime focus of the 4\,m Victor Blanco telescope at the Cerro Tololo Inter-American Observatory (CTIO) in Chile, as part of observing proposal 2017A-0911 (PI: Chies-Santos).  
Our initial goal was to make a mosaic of 4 DECam fields in order to uniformly cover a radius of $\sim 0.5$\,Mpc around NGC 3115, which would allow us to probe its system of satellites at distances equivalent to the outermost Galactic counterparts. Since we were only awarded half the requested time, we chose to observe 2 DECam pointings along the semi-major axis of NGC\,3115. 
DECam has an array of 59 scientific 2k × 4k CCD detectors with a 2.9 sq. deg field of view and a pixel scale of 0.27" (unbinned). These observations allowed us to reach distances around 200 times the effective radius of NGC 3115 ($R_e = 1.64 \ \rm kpc$) or $\sim 4$ times its virial radius\footnote{The effective radius is based on \citet{brodie14} and the virial radius is based on \citet{kravtsov13}}. We obtained a series of 6 $ \times $ 300 s dithered exposures in $g$ and $r$ bands under photometric conditions for 2 pointings.

\begin{figure*}
    \centering
    \includegraphics[width=0.85\linewidth]{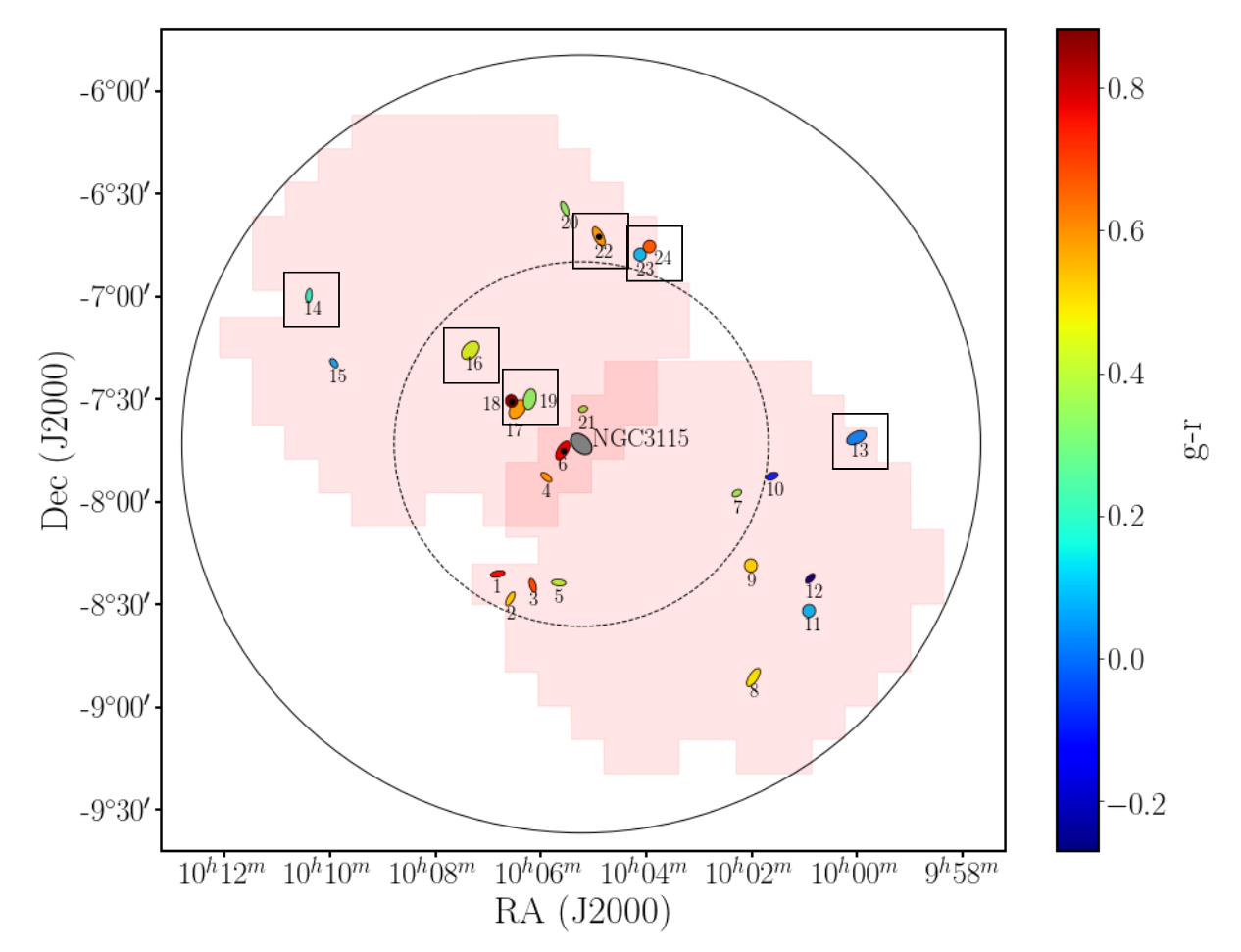}
    \caption{The spatial distribution of the LSBD candidates around NGC3115 (grey ellipse) overlayed on the two observed DECam observed pointings. North is up and East is to the left. The ellipticities, effective radius and position angles of the symbols were scaled by the values from Tab. \ref{tab:structure}. The nucleated candidates are indicated by a black dot. The dashed and solid circles have radii 150 kpc and 320 kpc, respectively. The symbol colours vary with the $g$-$r$ colour index, as shown by the vertical scale on the right. The GMOS pointings on the LSBds that were followed-up are indicated as the black squares.}
    \label{fig:observations}
\end{figure*}

In Fig.~\ref{fig:observations} we show the location on the sky of the two observed fields around NGC\,3115. The fully reduced and stacked images were produced by the DECam community pipeline \citep{valdes14}. In Tab.~\ref{tab:obs} the total exposure times, median seeing and the community pipeline zeropoints for the stacked final NE and SW pointings are outlined.
We checked and found good agreement between APASS \citep{henden15} and our  magnitudes with the community pipeline zeropoints applied (for more details on the photometry, see Sec. \ref{sec:analysis}). We therefore use the zeropoints from Tab.~\ref{tab:obs} to calibrate our data. The estimated image depth at SNR = 3 is 25.08 mag in the r band and 24.76 mag in the g band (considering the pointing with lower exposure time). To the best of our knowledge, these are the deepest data that exist at such large radii around NGC3115.

\begin{table}
    \centering
    \begin{tabular}{c|c|c|c|c|c|c|c}
  \hline
 & \multicolumn{3}{c}{$g$} & & \multicolumn{3}{c}{$r$} \\ 
  \cline{2-4}
  \cline{6-8} 
Pointing  & $t_{\rm{exp}}$ & $m_0$ & seeing  & &$t_{\rm{exp}}$ & $m_0$ & seeing \\
  & (s) & (mag) &(")   & & (s) & (mag) &(")  \\
\hline
 NE & 12599 &  31.012 &  1.45 & & 12599 & 31.395 &1.13\\
 SW & 12899 &  31.034 & 1.22&  &  12599 &  31.395 &1.13\\
    \end{tabular}
    \caption{Blanco/DECam Journal of Observations: The data used in this work were observed during 4 nights (February 15-18th, 2017). In the first couple of nights we observed the SW pointing and during the last couple of nights, the NE pointing. The total exposure time is $t_{\rm{exp}}$ and $m_0$ is the zero-point from the final stacked image as provided by the community pipeline. The seeing corresponds to the median FWHM measured by the same pipeline.}
    \label{tab:obs}
\end{table}

\begin{table}
    \centering
    \begin{tabular}{c|c|c|c|c|c|c|c}
  \hline
 & \multicolumn{3}{c}{$g$} & & \multicolumn{3}{c}{$i/z$} \\ 
  \cline{2-4}
  \cline{6-8} 
ID  & $t_{\rm{exp}}$ & $m_0$ & seeing  & &$t_{\rm{exp}}$ & $m_0$ & seeing \\
  & (s) & (mag) &(")   & & (s) & (mag) &(")\\
\hline
 13* & 3000 &   32.756 &  0.59 & & 3360 & 32.004 & 0.58\\
 14\dag        & 3000 &  32.881 & 0.64  &  &  1800 &  32.627 & 0.56\\
 16*        & 1800 &  32.503 & 0.66 &  &  1200 &  32.729 & 0.53\\
 17-19* & 1800 &  32.483 & 0.76  &  &  1200 &  32.884 & 0.54\\
 22\dag         & 6600 &  32.035 & 0.80  &  &  4950 & 31.836 & 0.57\\
 23-24\dag    & 3000 &  32.192 & 0.64 &  &  1800 &  32.273 & 0.52\\
    \end{tabular}
    \caption{Gemini Journal of Observations: The data used in this work were observed during 7 nights (November 27th to December 25th, 2020). The total exposure time is $t_{\rm{exp}}$ and $m_0$ is the zero-point from the final stacked image. The candidates observed in z-band are show in * and the candidates observed in i-band are shown in \dag. The seeing corresponds to the median FWHM measured using IRAF.}
    \label{tab:obs_gem}
\end{table}

To compare directly with photometry from other systems available from the literature, we employed the conversion method outlined by Lupton (2005), with the RMS photometric calibration error for $g$-$r$ determined to be 2\%  \citep{sdss05}, to transform our g and r magnitudes into B and V: \footnote{http://www.sdss3.org/dr10/algorithms/sdssUBVRITransform.php, where we take into account the conversion from SDSS $g$ and $r$ magnitudes which are close to AB $g$ and $r$.}

\begin{equation}
  V =g - 0.5784(g-r) - 0.0038,
 \label{eq:lupton_V}
\end{equation}

\begin{equation}
  B =g + 0.3130(g - r) + 0.2271.
 \label{eq:lupton_B}
\end{equation}

\subsection{Gemini/GMOS imaging}
\label{sec:gemini}
We followed up with GMOS/GEMINI 6 interesting fields that contained 9 dwarf galaxy candidates (GN-2020B-Q-223 e GS-2020B-Q-237, PI: Furlanetto). The data were observed during 7 nights (from November 27th to December 25th, 2020) in both Gemini South and Gemini North. In Tab. \ref{tab:obs_gem} we show the total exposure time, the magnitude zero-point and seeing for each field observed in Gemini and the LSBDs present in each field, the zero point magnitude were obtained using approximate calibrations from Pan-STARRS and Gaia. We obtained a series of 6x300s exposures in g band, 8x150s exposures in z band on Gemini North.
The images were reduced using the DRAGONS - GMOS data reduction tasks \citep{labrie2019}. 
The reduction steps consisted in co-adding different exposures, bias subtraction, flat-fielding, fringe removal on z band images, generating a bias and flat corrected files and stacking them.

\section{LSB Object Detection}
\label{sec:visua_inspect}
Our goal is to find new LSB dwarf galaxy candidates around NGC\,3115 as well as identify any known LSB dwarf objects that will appear diffuse in our DECam images. 
Because one can easily identify the candidates by considering simultaneously their sizes, colours and diffuse morphologies that arise from shallower surface brightness profiles (when compared to other galaxies) they are readily visible in the images. Moreover, two DECam fields are feasible to be visually inspected.
Thus, we follow an approach based on visual inspection [as previous works in the field, such as \citealt{eigenthaler18} and \citealt{muller17}] of the $g$ and $r$-band images to search for diffuse  dwarf galaxy candidates. For recent works that follow automated methods see \citep[e.g.][]{zaritsky19, prole19, Tanoglidis20}.

We created stamps to divide the work of the visual inspection. We generated 2000 stamps from the two observed fields. Each stamp image is $\sim 800\times 800$ pixel with 40 pixels of superposition (to avoid missing objects in the edge of the stamp). The stamp size is ($ \sim 3.6 $ arcmin $ \times $ 3.6 arcmin) and it corresponds to approximately 10 kpc at the distance of NGC3115. This is a suitable compromise between visual inspection of the smaller candidates and still enough to accommodate possible UDGs.

We developed and followed a tailored visual inspection tool to search for diffuse object candidates. The tool provides the stamps with an interface to flag candidates and possible image artifacts with just mouse clicks on the image. 
Six collaborators performed the visual inspection independently. The inspectors were divided into two groups (ACS, BXS and CF; WS, KA and AP), with each group inspecting one of the DECam tiles. For consistency, we analysed the two tiles independently, that is, in the small overlap region (see Fig. \ref{fig:observations}) between the two pointings, we did not stack them. Rather, we analysed them separately.

Candidates were selected when at least two inspectors (out of three of the group that inspected the image) clicked on the candidate within a maximum of $3$ arcsec distance. After cross-matching the list of identified objects of each inspector, we obtained a total of 40 candidate LSB dwarf galaxies.
Based on their surface brightness ($\mu_g > 22.5$ mag~arcsec$^{-2}$), structural parameters (see details in Sec. \ref{sec:profile_fitting}) and another iteration of visual inspection, we removed artifacts objects. Finally, we are left with 24 LSB dwarf candidates, whose distribution on the sky is shown in Fig.~\ref{fig:observations}.
In Tab. \ref{tab:names_coord} we present the positions and, for simplicity, our adopted ID naming used throughout the paper. 

\begin{table}
    \centering
    \begin{tabular}{lccc
    }
    \hline
    ID & Name & RA (J2000) & Dec (J2000) \\
    \hline
 1 & dw100649-082105 & 10:06:49.155 & -08:21:05.24  \\
2  & dw100634-082819 & 10:06:34.818 & -08:28:19.13 \\
3  & dw1006009-082426 & 10:06:09.229  & -08:24:26.36 \\
4  & dw100553-075248 & 10:05:53.759  & -07:52:48.97 \\
5  & dw100539-082337 & 10:05:39.688 & -08:23:37.28 \\
6*  & dw100535-074459 & 10:05:35.057  & -07:44:59.31 \\
7  & dw100216-075729 & 10:02:16.977 & -07:57:29.82 \\
8  & dw100158-085117 & 10:01:58.141 & -08:51:17.70 \\
9*  & dw100201-081836 & 10:02:01.053   & -08:18:36.60 \\
10 & dw100136-075230 & 10:01:36.995  & -07:52:30.49 \\
11* & dw100054-083149 & 10:00:54.926  & -08:31:49.70 \\
12 & dw100053-082223 & 10:00:53.542  & -08:22:23.22 \\
13* & dw100000-074116 & 10:00:00.851 & -07:41:16.90 \\
14 & dw101023-065948 & 10:10:23.914 & -06:59:48.31 \\
15 & dw100955-071929 & 10:09:55.493  & -07:19:29.69 \\
16* & dw100720-071547 & 10:07:20.228 & -07:15:47.51 \\
17* & dw100626-073257 & 10:06:26.642 & -07:32:57.41 \\
18* & dw100633-073033 & 10:06:33.600 & -07:30:33.05 \\
19* & dw100612-073002 & 10:06:12.626 & -07:30:02.41 \\
20 & dw100532-063420 & 10:05:32.809  & -06:34:20.37 \\
21 & dw100512-073256 & 10:05:12.004  & -07:32:56.75 \\
22 & dw100454-064231 & 10:04:54.206 & -06:42:31.69 \\
23 & dw100407-064747 & 10:04:07.131   & -06:47:47.06 \\
24 & dw100356-064527 & 10:03:56.460  & -06:45:27.86
 
    \end{tabular}
    \caption{ID and equatorial coordinates of the LSB dwarf candidates found in this work. The objects in common with \citet{carlsten21b} are flagged with a *.}
    \label{tab:names_coord}
\end{table}

Some of our candidate objects have already been reported in the literature. 
\citet{cantiello18} reported eight LSB galaxies around NGC\,3115 and five are in common with ours (Id 4, 17, 18, 19, 21). Object Id 6 has been previously studied by \citet{sharina05} and \citet{puzia08} and is also known as KK84.
While finalizing this manuscript, the works of \citet{carlsten21a, carlsten21b} came out. They report 12 and 14 LSBds around NGC\,3115 respectively. Eight objects are present in the \citet{Carlsten22} sample, all are in common with the sample we report here, they are flagged in Table \ref{tab:names_coord}. Moreover, they also identified 3 more dwarfs that are not in our field. They also found two LSBd candidate that are in our field-of-view, but we did not find it. This is probably because it is very near a bright star and the star spikes make it hard to identify the dwarf candidate through our applied visual inspection methodology.

\section{LSB Dwarf Candidate properties}
\label{sec:analysis}
In Sec.\,\ref{sec:profile_fitting} we present the surface photometry and structural analysis for the LSB dwarf candidates selected in Sec.\,\ref{sec:visua_inspect} and in Sec. \ref{sec:prop} we explore the global properties of the sample of objects and compare them to similar samples from the literature. 

\subsection{Surface  Photometry \& Profile Fitting}\label{sec:profile_fitting}

For each candidate we produce a 400 pixel $\times$ 400 pixel ($\sim 108 $ arcsec $ \times 108 $ arcsec) stamp from the $g$- and $r$-band stacked images. 
We use {\sc SExtractor} \citep{bertin96} to generate segmentation maps for each stamp using 2$\sigma$ threshold above background and mask all non-LSB galaxy sources. We also use {\sc SExtractor} to estimate the sky background locally in each stamp. The parameters used for background estimation and segmentation are presented in Tab. \ref{tab:sextractor}.

\begin{table}
\centering
\caption{SExtractor input parameters used in the LSB dwarf candidates stamps for the estimate of background and segmentation.}
\label{tab:sextractor}
\begin{tabular}{cc} \hline
Parameter name & Input Configuration \\ \hline
DETECT\_MINAREA & 4 pixels \\
DETECT\_THRESH & 2 \\
ANALYSIS\_THRESH & 2 \\
DEBLEND\_NTHRESH & 32 \\
DEBLEND\_MINCONT & 0.005 \\
BACK\_SIZE & 64 pixels \\
BACKPHOTO\_TYPE & GLOBAL
\end{tabular}
\end{table}

We proceed by modelling the point spread function (PSF) for each galaxy candidate using a 2D Gaussian profile with FWHM equal to the seeing of the correspondent pointing (see Tab. \ref{tab:obs}). We check that the variation of PSF FWHM in the surrounding areas of our LSB dwarf  candidates with respect to the median FWHM measured by the DECam pipeline is very small, having negligible effect on the derived photometric properties. 

As the surface brightness profiles of dwarf galaxies are usually well described by the Sérsic model \citep{sersic68}, all candidate LSB dwarfs in this work were fitted using 2D ellipsoidal single-component Sérsic profiles. We perform profile fitting separately for each band using {\sc imfit}\footnote{\url{http://www.mpe.mpg.de/~erwin/code/imfit/}} \citep{erwin2015}, a program for fitting astronomical images, especially images of galaxies. The one-dimensional S\'ersic profile is described by the following equation:
\begin{equation}
\centering
I(R)=I_e \exp\left\{ -b_n\left[ \left( \frac{R}{R_e}\right) ^{1/n} -1\right] \right\},
\end{equation}
where $I_e$ is the intensity at the effective radius $R_e$ that encloses half of the total light from the model and $n$ is the Sérsic index, which describes the profile shape  \citep{graham2005}. The constant $b_n$ is defined in terms of the parameter $n$ and is computed using the polynomial approximation of \citet{Ciotti99} when $n > 0.36$ and the approximation of \citet{MacArthur03} when $n \leq 0.36$. Here we fit two-dimensional profiles, so $R$ is the radial distance along the major-axis and the models also have 2 geometric parameters, the ellipticity ($\epsilon$) and the position angle (PA).

To increase the stability of the fitting process, we model a uniform sky background alongside the Sérsic component and the intensity of the background is fixed to the value obtained by {\sc SExtractor}. The profile fitting was done using Gaussian PSFs and weight maps. For the nucleated candidates, which have a bright point source within 1\arcsec of the centre of the galaxy, it was necessary to mask the nuclei in order to fit well their overall profile and avoid an overestimation of the Sérsic index. 

Our fitting process is divided into three distinct steps to ensure robust results. In the first step, we initiate the process by selecting initial estimates for parameters such as $R_e$, PA (position angle), and $\epsilon$ (ellipticity) based on visual inspection. We set the Sérsic index ($n$) to 1 and allow all parameters to freely vary during the minimization process. For this initial step, we employ the Levenberg-Marquardt algorithm. 

Moving on to the second step, we refine the best-fit models for objects where the initial fit did not converge or where there were noticeable discrepancies between the model and data residuals (e.g., incorrect PA, over-subtraction, etc.). This step involves an iterative approach: we keep one parameter fixed (typically PA, $\epsilon$, or $n$) while providing different initial estimates for the other parameters. Following the minimization with the fixed parameter, we then allow the previously fixed parameter to vary freely while fixing the others. This process iterates until a significantly improved fit is achieved. Model acceptance or rejection is determined based on the reduced $\chi^2$ statistic and a thorough examination of residual images.

The third and final step is designed to test the stability of the best-fit models. To do this, we repeat the fitting procedure using the Differential Evolution (DE) minimization algorithm \citep{storn97} implemented in {\sc imfit}. This algorithm is less prone to getting trapped in local minima of the $\chi^2$ landscape, as discussed in \citet{erwin2015}. Notably, the DE algorithm does not require initial estimates; instead, it relies on parameter intervals for estimation. For all galaxies in this step, we consider the following parameter intervals: PA $= [0,360]$ degrees, $\epsilon = [0,1]$, $n = [0.2,8]$, and $R_e = [0.1,54]$ arcseconds.

After this, we compare the result obtained in this step to the one found in the second step. The final best-fit parameters resulting from this approach are presented in Tab. \ref{tab:structure}, and model images as well as residuals are detailed in Appendix \ref{sec:mock}.

We obtain the total apparent magnitudes and surface brightness in $g$- and $r$-bands using the parameters of the Sérsic model, following \cite{graham2005}. 
The total flux is computed as:
\begin{equation}
    F = 2\pi R_e^2 I_e \frac{n e^{b_n}}{b_n^{2n}} \Gamma(2n) (1-\epsilon),
\end{equation}
where $\Gamma(x)$ is the gamma function and the geometrical correction (1-$\epsilon$) is introduced to take into account the ellipticity of the model. The mean surface brightness within one effective radius, $\langle \mu \rangle_e$,  is computed using the following equation from \cite{graham2005}: 
\begin{equation}
    \langle \mu \rangle_e = m + 2.5\log(2\pi R_e^2),
\end{equation}
where $m$ is the magnitude computed using $F$ and the zeropoint magnitudes from Tab. \ref{tab:obs}. We adopt Galactic extinctions from \cite{schlafly11}: 0.101 mag for the $r$-band and 0.150 mag for the $g$-band. In Tab.~\ref{tab:photometry} we show the $g$- and $r$-band magnitudes and the surface brightness for each candidate. 
The reliability of the profile fitting process and surface photometry is addressed in Appendix \ref{sec:mock}, where we show deviations for fitted parameters using mock galaxies injected in our stacked images. We use the same sample of mock galaxies to estimate  the uncertainties of the structural and photometric quantities presented in Tables \ref{tab:structure} and \ref{tab:photometry}.

\subsection{Global properties of the LSB candidates}
\label{sec:prop}

Here we explore the global properties of our LSB dwarf candidates and compare them to previous works.
We note that the LSB dwarf candidate ID6 was previously studied by \cite{sharina05}. While they report $M_V=-14.4$ we obtain $M_V=-14.32$ by applying Eq.\ref{eq:lupton_V}. As for the surface brightness, they report $\mu_B=25.4$mag arcsec$^{-2}$. Adopting the $g$-band measured $R_e$ and converting our magnitudes to the B band we find $\mu_B=24.94$mag arcsec$^{-2}$.
While \cite{cantiello18} reports objects ID4, ID17, ID18, ID19 and ID21, they do not provide detailed properties.

\begin{figure}
    \centering
    \includegraphics[width=1.0\linewidth]{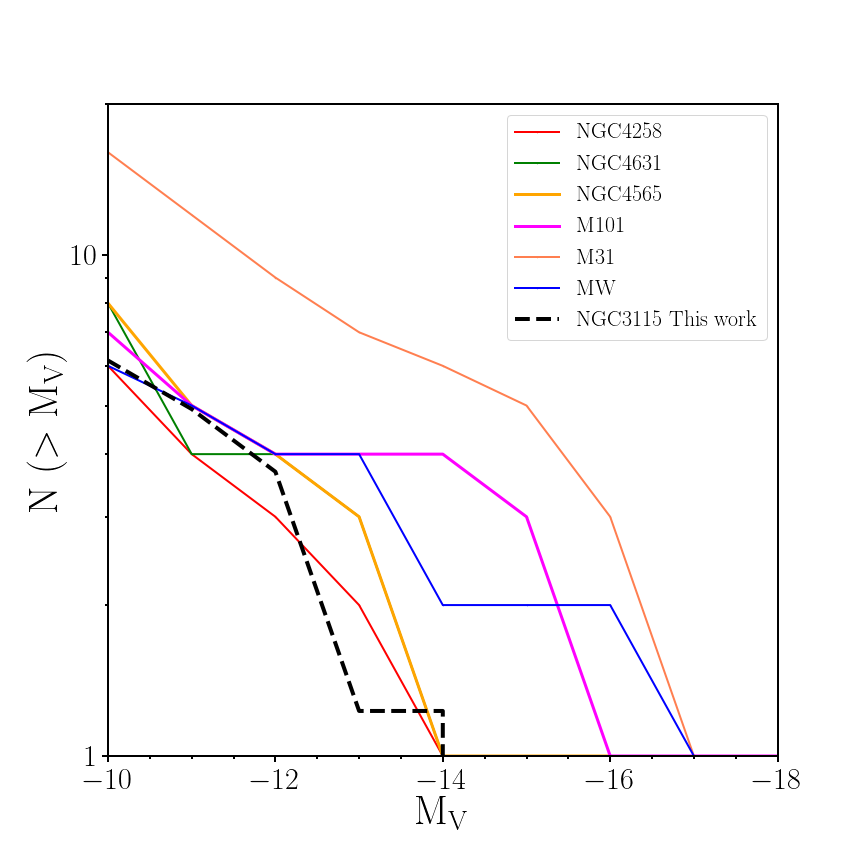}
    \caption{Cumulative luminosity function (completeness-corrected) and comparison with \citet{carlsten2020b} sample. The x-axis represents the magnitude in the V band in AB.}
    \label{fig:luminosity_function}
\end{figure}

In Fig.~\ref{fig:luminosity_function} we compare the luminosity functions of our candidate galaxies with those of the \citet{carlsten2020b} Local Volume sample, from which it follows that the luminosity function of the NGC\,3115 satellite system is comparable to those of NGC 4258 and NGC 4565.
Following \citet{carlsten2020b}, the expected number of satellites with $M_V<-9$ for a projected radius of $< 150 \ \rm kpc$ from the centre of a host with a stellar mass similar to that of NGC3115 (log M$_{\star}$/M$_{\odot}$=10.93; \citealt{alabi2017}) should be $\sim$ 12 objects. By extrapolating the anticipated values from \citet{carlsten2020b} and counting the objects within the projected radius for NGC 3115 (as indicated by the inner radius in Fig.~\ref{fig:observations}), we estimate a total of 7 satellites. This is $\sim58\%$ of the expected number in an area that encompasses 81\% of such projected radius, which is within the uncertainties provided by \citet{carlsten2020b}.

\begin{figure}
    \centering
    \includegraphics[width=1.0\linewidth]{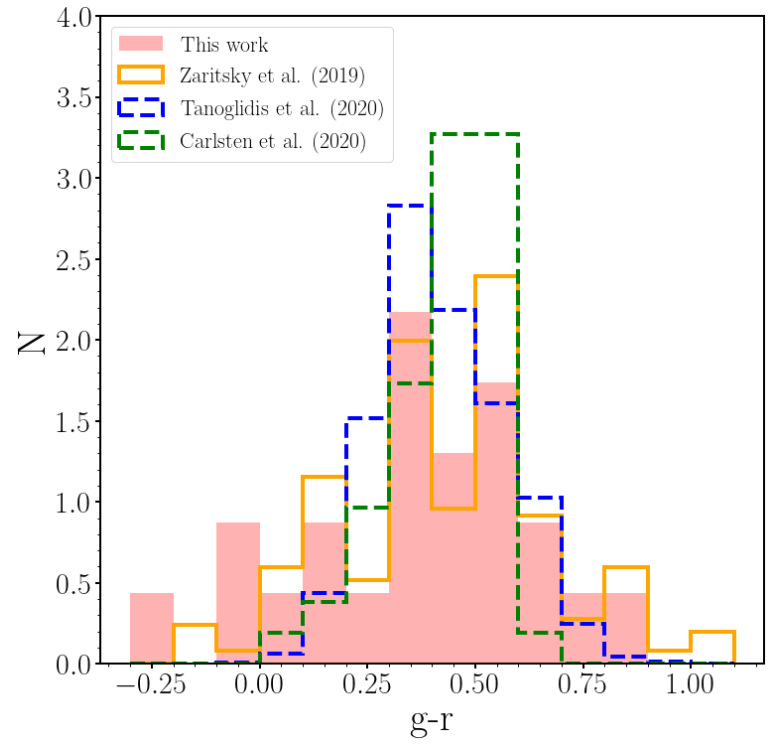}
    \caption{The ($g$-$r$) colour distribution for our candidate LSB dwarf in the filled histogram. DES \citep{Tanoglidis20}, Coma Cluster UDGs \citep{zaritsky19} and Local Volume sample \citep{carlsten2020a} are shown as the blue dashed line and green dashed line histograms respectively. The datasets have been normalized to the same scale.}
    \label{fig:colour_hist}
\end{figure}

 In Fig. ~\ref{fig:colour_hist}, we illustrate a comparison of the colour distribution of our candidate objects with two reference datasets: the Local Volume sample from \cite{carlsten2020b} and the DES sample described in \cite{Tanoglidis20}. 
 We found that our sample displays a relatively broad range in colour roughly similar to that found in \cite{carlsten2020b} and \cite{Tanoglidis20}. However, within our sample, there are objects with colours that are both redder, as well as galaxies that appear bluer compared to those in other studies. \cite{md2016,roman2019, prole19a, prole19, Martin19}, as shown in Fig \ref{fig:colour_hist}.
 Our candidate LSB dwarf galaxies have median, mean and standard deviation ($g$-$r$) colours $0.40$, $0.38$ and $0.28$ respectively.
 From Fig.~\ref{fig:observations}, it is apparent that LSB dwarf colours are related to the separation with respect to NGC\,3115. While bluer LSB dwarfs tend to be located farther away from the main galaxy, redder LSB dwarfs are found closer to the centre. This suggest that the interaction with the environment can reduce the star formation of those closer to the centre  \citep{Greco18}.
 
\begin{table*}
\centering
\caption{Structural properties of the galaxies obtained for $g$ and $r$ bands on DECam. \textit{Type: }  \textbigcircle \ are non-nucleated, $\odot$ are nucleated, \textit{R$_e$: } effective radius, \textit{n: } Sérsic index, \textit{$\epsilon$: } ellipticity value.  }
\label{tab:structure}
\begin{tabular}{lllllllll}
\hline
&  & \multicolumn{3}{c}{$g$} && \multicolumn{3}{c}{$r$} \\ 
\cline{3-5}
\cline{7-9}
ID& Type & $R_e$ & n & $\epsilon$& & $R_e$ & n & $\epsilon$ \\
 &  & (pc) &  &  && (pc) &  &  \\ \hline
1  & \textbigcircle & 91.27 $\pm$ 63.80            & 0.71 $\pm$ 0.88             & 0.12 $\pm$ 0.26 &  & 103.58 $\pm$ 20.00           & 0.75 $\pm$ 0.23         & 0.12 $\pm$ 0.10   \\
2  & \textbigcircle & 66.06 $\pm$ 12.74 & 0.46 $\pm$ 0.30                        & 0.08 $\pm$ 0.11 & & 150.79 $\pm$ 46.83            & 1.29 $\pm$ 0.29         & 0.06 $\pm$ 0.07  \\
3  & \textbigcircle & 108.93 $\pm$ 98.40           & 0.59 $\pm$ 1.00             & 0.15 $\pm$ 0.49 &  & 111.63 $\pm$ 70.97           & 0.98 $\pm$ 0.59         & 0.15 $\pm$ 0.19   \\
4  & \textbigcircle & 89.95 $\pm$ 29.18            & 0.81 $\pm$ 0.52             & 0.21 $\pm$ 0.16 &  & 124.70 $\pm$ 47.87           & 1.38 $\pm$ 0.48         & 0.20 $\pm$ 0.08 \\
5  & \textbigcircle & 63.61 $\pm$ 31.68           & 1.00 $\pm$ 0.39           & 0.15 $\pm$ 0.11 & & 68.19 $\pm$ 30.27             & 1.32 $\pm$ 0.57         & 0.15 $\pm$ 0.05  \\
6  & $\odot$                       & 875.11 $\pm$ 78.32           & 0.87 $\pm$ 0.14             & 0.39 $\pm$ 0.01 &  & 1071.88 $\pm$ 472.87           & 0.68 $\pm$ 0.28         & 0.39 $\pm$ 0.01 \\
7  & \textbigcircle & 096.21 $\pm$ 27.45           & 0.90 $\pm$ 0.31              & 0.39 $\pm$ 0.06 & & 92.38 $\pm$ 59.15             & 0.71 $\pm$ 0.20         & 0.35 $\pm$ 0.03 \\
8  & \textbigcircle & 438.55 $\pm$ 74.31           & 0.41 $\pm$ 0.45 & 0.48 $\pm$ 0.14 &  & 397.28 $\pm$ 74.41                       & 0.61 $\pm$ 0.21         & 0.49 $\pm$ 0.07  \\
9  & \textbigcircle & 226.73 $\pm$ 69.50           & 0.82 $\pm$ 0.32             & 0.21 $\pm$ 0.10 & & 270.71 $\pm$ 26.14            & 0.37 $\pm$ 0.10          & 0.49 $\pm$ 0.01  \\
10 & \textbigcircle & 364.97 $\pm$ 148.93          & 0.66 $\pm$ 0.50             & 0.58 $\pm$ 0.06 & & 265.10 $\pm$ 68.02            & 0.48 $\pm$ 0.48         & 0.51 $\pm$ 0.04  \\
11 & \textbigcircle & 257.13 $\pm$ 68.98            & 0.84 $\pm$ 0.35               & 0.22 $\pm$ 0.17 &  & 254.22 $\pm$ 112.32       & 0.94 $\pm$ 0.36        & 0.22 $\pm$ 0.09  \\
12 & \textbigcircle & 156.50 $\pm$ 38.01            & 1.07 $\pm$ 0.27               & 0.29 $\pm$ 0.08 & & 175.54 $\pm$ 42.47           & 1.10 $\pm$ 0.27      & 0.26 $\pm$ 0.05  \\
13 & \textbigcircle & 612.62 $\pm$ 66.10           & 0.60 $\pm$ 0.12 & 0.30 $\pm$ 0.16   &  & 562.86 $\pm$ 26.58           & 0.73 $\pm$ 0.10 & 0.25 $\pm$ 0.01 \\
14 & \textbigcircle & 154.44 $\pm$ 49.63           & 1.27 $\pm$ 0.50              & 0.17 $\pm$ 0.23 & & 118.87 $\pm$ 25.61           & 0.93 $\pm$ 0.25       & 0.05 $\pm$ 0.12  \\
15 & \textbigcircle & 158.14 $\pm$ 49.66            & 0.70 $\pm$ 0.39             & 0.37 $\pm$ 0.16 & & 141.07 $\pm$ 41.81           & 0.72 $\pm$ 0.26         & 0.22 $\pm$ 0.15  \\
16 & \textbigcircle & 551.02 $\pm$ 35.67           & 0.90 $\pm$ 0.08             & 0.14 $\pm$ 0.02 & & 664.31 $\pm$ 49.70             & 0.88 $\pm$ 0.03          & 0.14 $\pm$ 0.01  \\
17 & \textbigcircle & 452.37 $\pm$ 47.53 & 0.75 $\pm$ 0.12             & 0.18 $\pm$ 0.07 & & 504.26 $\pm$ 48.84           & 0.92 $\pm$ 0.13              & 0.16 $\pm$ 0.04 \\
18 & $\odot$                       & 295.40 $\pm$ 85.38 & 0.68 $\pm$ 0.27               & 0.31 $\pm$ 0.10 &  & 409.37 $\pm$ 58.46           & 1.05 $\pm$ 0.14               & 0.30 $\pm$ 0.05   \\
19 & \textbigcircle & 422.76 $\pm$ 40.91           & 0.90 $\pm$ 0.14             & 0.26 $\pm$ 0.03 & & 384.20 $\pm$ 21.88 & 0.81 $\pm$ 0.03              & 0.24 $\pm$ 0.01  \\
20 & \textbigcircle & 190.79 $\pm$ 70.87           & 0.67 $\pm$ 0.44             & 0.08 $\pm$ 0.46 & & 235.96 $\pm$ 96.24            & 1.24 $\pm$ 0.36      & 0.21 $\pm$ 0.15   \\
21 & \textbigcircle & 106.82 $\pm$ 44.26           & 0.14 $\pm$ 0.59              & 0.46 $\pm$ 0.16 &  & 111.50 $\pm$ 38.41          & 0.29 $\pm$ 0.43      & 0.44 $\pm$ 0.10  \\
22 & $\odot$                       & 402.94 $\pm$ 43.91           & 0.67 $\pm$ 0.11              & 0.44 $\pm$ 0.04 & & 422.73 $\pm$ 33.99           & 0.72 $\pm$ 0.05      & 0.41 $\pm$ 0.02  \\
23 & \textbigcircle & 203.51 $\pm$ 63.41           & 1.08 $\pm$ 0.36              & 0.26 $\pm$ 0.10 & & 164.47 $\pm$ 51.01           & 0.77 $\pm$ 0.26      & 0.26 $\pm$ 0.05  \\
24 & \textbigcircle & 273.42 $\pm$ 39.87           & 1.90 $\pm$ 0.42              & 0.27 $\pm$ 0.30  &  & 264.66 $\pm$ 57.10           & 1.76 $\pm$ 0.28     & 0.19 $\pm$ 0.13  

\end{tabular}
\end{table*}

\begin{table*}
\caption{Photometric properties of the galaxies. Assumed distance modulus $(m-M) = 29.93 \pm 0.09$ mag \citep{Tonry01}. The uncertainties in the quantities were estimated as described in Appendix \ref{sec:mock}.}
\label{tab:photometry}
    \centering
    \begin{tabular}{|l|l|l|l|l|l|l|l|}
    \hline
    \multicolumn{3}{c}{$g$} & & \multicolumn{3}{c}{$r$} \\ 
  \cline{2-4}
  \cline{6-8} 

  ID & $m$ & $M$ & $\langle \mu \rangle_e$&  & $m$& $M$ &  $\langle \mu \rangle_e$ \\
 & (mag) & (mag) &  (mag\,arcsec$^{-2}$) & & (mag) & (mag) & (mag\,arcsec$^{-2}$) \\ \hline
1  & 22.06$\pm$0.55 & -07.87$\pm$0.56  & 25.65                           &  & 21.31$\pm$0.12 & -08.62$\pm$0.15  & 25.12                         \\
2  & 21.33$\pm$0.17 & -08.60$\pm$0.19   & 24.21                          &  & 20.82$\pm$0.25 & -09.11$\pm$0.28  & 24.85                         \\
3  & 22.70$\pm$0.89  & -07.23$\pm$0.89  & 26.67                          &  & 22.02$\pm$0.59 & -07.91$\pm$0.60   & 25.90                           \\
4  & 21.57$\pm$0.34 & -08.36$\pm$0.35  & 25.13                          &  & 21.02$\pm$0.32 & -08.91$\pm$0.33  & 25.14                         \\
5  & 21.16$\pm$0.36 & -08.77$\pm$0.37  & 23.96                          &  & 20.78$\pm$0.21 & -09.15$\pm$0.23  & 23.66                           \\
6  & 15.95$\pm$0.09 & -13.98$\pm$0.13 & 24.50                          &  & 15.16$\pm$0.55 & -14.77$\pm$0.56 & 24.05                         \\
7  & 20.79$\pm$0.22 & -09.14$\pm$0.24  & 24.49                           &  & 20.45$\pm$0.20  & -09.48$\pm$0.22  & 24.01                          \\
8  & 20.77$\pm$0.47 & -09.16$\pm$0.48  & 27.76                          &  & 20.29$\pm$0.47 & -09.64$\pm$0.48  & 27.20                          \\
9  & 20.09$\pm$0.34 & -09.84$\pm$0.35  & 25.65                          &  & 19.59$\pm$0.01 & -10.34$\pm$0.13 & 25.29                          \\
10 & 20.30$\pm$0.90   & -09.63$\pm$0.91  & 26.90                          &  & 20.33$\pm$0.51 & -09.54$\pm$0.52  & 26.22                          \\
11 & 20.37$\pm$0.29 & -09.56$\pm$0.30   & 26.20                          &  & 20.27$\pm$0.50  & -09.66$\pm$0.50   & 26.04                          \\
12 & 20.57$\pm$0.17 & -09.36$\pm$0.20   & 25.33                          &  & 20.84$\pm$0.25 & -09.09$\pm$0.27  & 25.23                           \\
13 & 17.00$\pm$0.15    & -12.93$\pm$0.17 & 24.72                          &  & 16.96$\pm$0.04 & -12.97$\pm$0.10  & 24.44                         \\
14 & 21.09$\pm$0.29 & -08.84$\pm$0.30   & 25.82                         &  & 20.91$\pm$0.16 & -09.02$\pm$0.19  & 25.02                          \\
15 & 21.27$\pm$0.40  & -08.66$\pm$0.41  & 26.06                         &  & 21.20$\pm$0.32  & -08.73$\pm$0.33  & 25.76                          \\
16 & 17.39$\pm$0.07 & -12.54$\pm$0.11 & 24.88                          &  & 16.97$\pm$0.06 & -12.96$\pm$0.11 & 24.61                          \\
17 & 18.93$\pm$0.15 & -11.00$\pm$0.18    & 25.99                          &  & 18.36$\pm$0.15 & -11.57$\pm$0.17 & 25.61                          \\
18 & 20.05$\pm$0.37 & -09.88$\pm$0.38  & 26.19                           &  & 19.17$\pm$0.12 & -10.76$\pm$0.15 & 25.97                          \\
19 & 18.12$\pm$0.15 & -11.81$\pm$0.17 & 25.04                         &  & 17.80$\pm$0.05  & -12.13$\pm$0.10  & 24.43                           \\
20 & 21.63$\pm$0.48 & -08.30$\pm$0.49   & 26.82                          &  & 21.31$\pm$0.52 & -08.62$\pm$0.53  & 26.26                          \\
21 & 22.18$\pm$0.27 & -07.75$\pm$0.30   & 26.12                         &  & 21.83$\pm$0.27 & -08.10$\pm$0.28  & 25.80                          \\
22 & 19.17$\pm$0.09 & -10.75$\pm$0.13 & 26.00                          &  & 18.60$\pm$0.08  & -11.33$\pm$0.12 & 25.53                           \\
23 & 20.19$\pm$0.34 & -09.74$\pm$0.35  & 25.52                          &  & 20.09$\pm$0.25 & -09.84$\pm$0.27  & 24.91                         \\
24 & 21.18$\pm$0.20  & -08.75$\pm$0.22  & 27.15                          &  & 20.52$\pm$0.22 & -09.41$\pm$0.24  & 26.37 
    \end{tabular}
\end{table*}

\begin{figure*}
    \centering
    \includegraphics[width=0.9\linewidth]{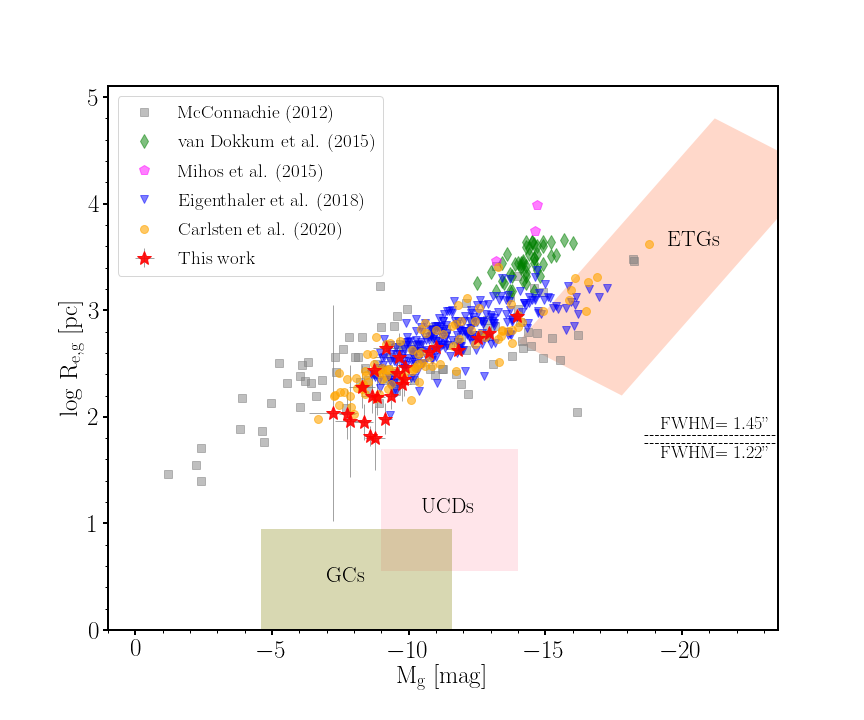}
    \caption{The magnitude (\(M_g\)) - size (effective radius in the g-band) diagram for our sample LSB dwarf galaxies as red stars. Literature LSB dwarf galaxies are also shown for comparison with symbols according to the legend. The expected loci of globular clusters (GCs), ultra-compact dwarf galaxies (UCDs), and early-type galaxies (ETGs) are also shown for completeness (from \protect\cite{eigenthaler18} and references therein).}

    \label{fig:money_plot}
\end{figure*}

In  Fig.~\ref{fig:money_plot} we show the $g$-band magnitude - size diagram for our sample LSB dwarf galaxies and compare to literature samples. 
Our LSBds fall in the same loci as the faint end of \cite{carlsten2020a} and \cite{eigenthaler18} sample, not covering the UDG Coma cluster sample from \cite{vD15}.

\section{Globular Cluster Candidates}\label{sec:GCs}
While inspecting the residual images of our candidate LSB dwarf galaxies (see Sec.\ref{sec:profile_fitting}) we noted that several of them present point compact sources (see Fig. \ref{fig:stamps}), that could be globular cluster candidates. We followed-up these LSB dwarf candidates with Gemini/GMOS.
The data reduction procedures for this imaging set are presented in Sec. \ref{sec:gemini}. 
In the following we outline the methodology we adopt to detect GC candidates around the LSB dwarfs ID13, ID14, ID16, ID17, ID18, ID19, ID22, ID 23 and ID 24 through Gemini/GMOS imaging. 

\begin{figure*}
    \centering
    \includegraphics[width=1\linewidth]{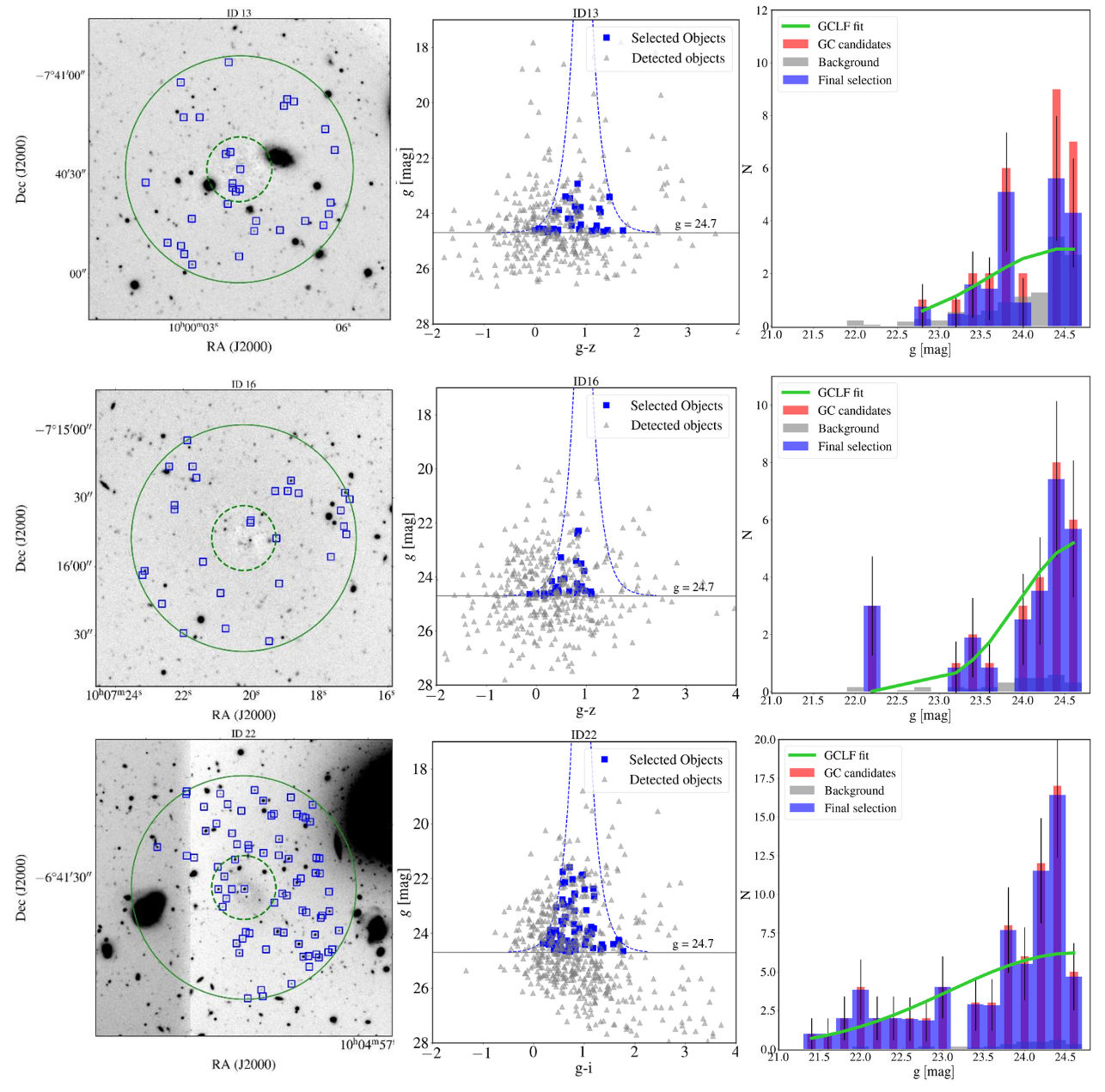}
    \caption{\textit{Left panel}: The spatial distribution of the GC candidates of LSBds 13, 16 and 22 sub-sampled on $MAG\_AUTO\_g \le 24.7$ marked in blue squares. The solid green line indicates the distance of 3.5 $R_{e}$ and the dashed green line the distance of 1  $R_{e}$.  \textit{Middle panel}: the colour-magnitude diagram (CMD) for the respective LSBds. The blue dashed lines indicate the boundary selection in colour and the horizontal line represents the cut in magnitude. \textit{Right panel}: Number of GCs per magnitude bin for the respective LSBds with the GCLF fit along. The grey bars are the background contamination, the red bars are the selection without background subtraction, the blue bars with errors are the final selection with background subtraction and the green line are the GCLF.}
    \label{fig:13dist}
\end{figure*}

Initially, we produce stamps with a size of 128.244 arcsec x 128.244 arcsec.
Then we follow the procedures adopted for DECam for surface photometry and profile fitting with \texttt{IMFIT} (see Sec. \ref{sec:profile_fitting}) and produce residual images. To measure the GC candidates, we employed aperture photometry. We proceed by running {\sc SExtractor} in dual-band image mode to identify the sources within the residual images of the LSBd candidates. In Tab. \ref{tab:source}, we provide the parameters utilized for detecting these sources. We set the detection images based on the better seeing band of the observation (see Tab. \ref{tab:obs_gem}).
As an example, in the left and middle panels of Fig. \ref{fig:13dist} we show the distribution on the sky and the $g-z$ vs. $g$ CMD for the selected sources around LSBd 13. 
Blue dots indicate the selected candidates with CLASS\_STAR\_g $\geq 0.05$.
Such CLASS\_STAR\ criterium is chosen in order to exclude galaxy subtraction artifacts.
Similar figures for all LSB dwarf candidates are presented in the Appendix \ref{sec:indivual_plots} (Fig. \ref{fig:gcs}). 

\begin{figure}
    \centering
    \includegraphics[scale=0.50]{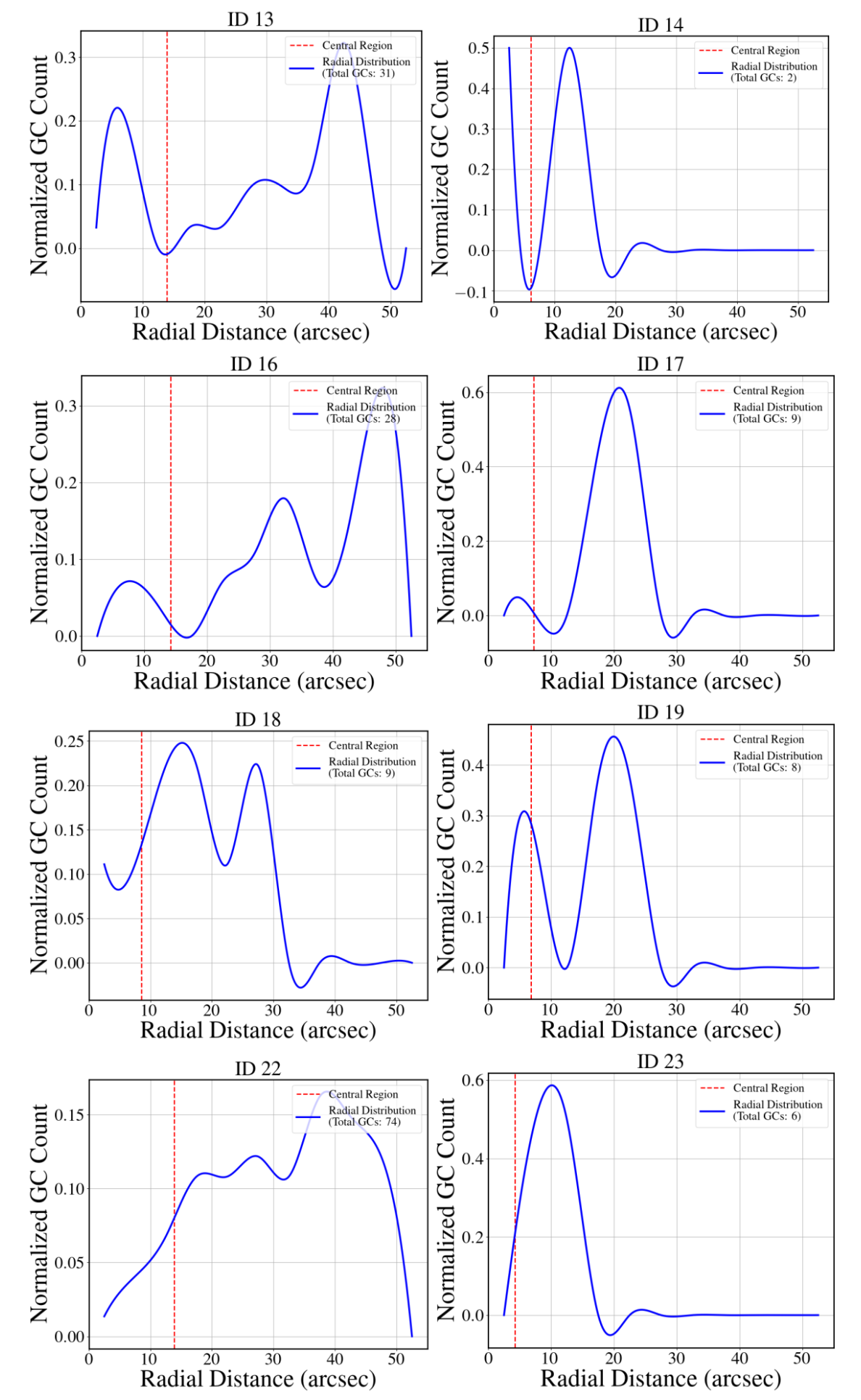}
    \caption{Radial distribution of the GC candidates, normalized by the GC count within each LSBd. The red dotted line indicates the distance of one effective radius (R$_{e}$). Each plot shows the total number of point sources selected as GC candidates.}
    \label{fig:radialDist}
\end{figure}

The selection of the globular cluster  sample was based on photometric errors, which exhibit an exponential-like shape with respect to magnitude (e.g. \citealt{exponential_errors}). Thus, we fit an exponential law encompassing the expected colour range for the NGC\,3115 GC system as described by \citealt{Faifer11, Forbes2017}. The fit is based around the median value of the confirmed GCs of NGC\,3115, until reaching 2 mag below the expected turnover magnitude at the distance of the host galaxy (see Fig. \ref{fig:bigCMDGCs}). 
In Fig. \ref{fig:radialDist}, we present the radial distribution of the objects identified as GC candidates. To generate the smooth lines we had applied the Spline interpolation method available on the \texttt{scipy} package. The figure reveals that certain LSBds exhibit a peak near the region of an R$_{e}$, suggesting the presence of a central clustering region in these galaxies.

\begin{table}
	\centering
	\caption{Source extraction photometry parameters used in order to identify our GCs candidates.
	}
	\label{tab:source}
	\begin{tabular}{lccr} 
		\hline
		Parameter name & Input configuration\\
		\hline
		DETECT\_MINAREA & 4\\
		DETECT\_THRESH  &  2\\
		ANALYSIS\_THRESH & 1.5\\
		DEBLEND\_NTHRESH & 64 \\
		DEBLEND\_MINCONT  & 0.005\\
		PHOT\_APERTURES  & 4, 5, 6, 7\\
		PHOT\_AUTOPARAMS & 4, 4.5 \\
		WEIGHT\_TYPE	& MAP\_WEIGHT\\
		
		\hline
	\end{tabular}
\end{table}

To account for background contamination we apply our selection method to a distant  annular region with 10 < R/R$_e$ < 16 of the host galaxies in each pointing obtained from GMOS.
This allowed us to estimate the number of objects misclassified as GCs due to their background status while minimising the probability of classifying actual GCs as background objects. Background objects were subsequently subtracted from the magnitude histograms (see Fig. \ref{fig:13dist}).

Various studies have shown that the number of GCs ($N_{GC}$) correlates with the total mass of the host galaxy 
(\citealt{Blakeslee1997, Spitler09, Hudson2014, harris2017, forbes18, zaritsky22}, to name a few). Here we estimate the total number of clusters on the LSB dwarfs around NGC\,3115 based on properties of the globular cluster luminosity function (GCLF).

To calculate the GCLF we assume a Gaussian distribution and used a Markov Chain Monte Carlo (MCMC) algorithm through the python package {\tt emcee} \citep{Foreman13} to obtain the posterior distributions of the three Gaussian g-band parameters ($\mu_g, \sigma_g, a_g$) for the GC systems of each LSBd. As a first guess, we used the parameters based on the distribution of GCs systems on LSBds.

In order to obtain the $N_{GC}$ we used the Python package {\tt scipy} with the function integrate.quad to integrate the Gaussian with the values to the g-band GCLF distribution obtained previously with MCMC. The number of GCs for each galaxy was calculated as: 
    \begin{equation}
         N_{GC} = \int^{\infty}_{-\infty} a\cdot e^{-\frac{(x-\mu)^2}{(2 \cdot \sigma^2)}}dx,
    \end{equation}
where $x$ represents the magnitude in which the GCLF was built.

In our study, we observe that some LSBds with a lower number of GC candidates, selected according to our criteria, exhibited models that did not fit well, such as galaxies ID17, ID18, ID19 and ID23. It is important to acknowledge that the performance of our MCMC model was limited due to the relatively low number of available GC candidates for analysis, which influenced the model accuracy. For those we have decided to count the number of GCs that remain after the background decontamination. 

We construct the GCLF based on the $g$-band magnitude, as this is the common band for all LSB dwarfs observed with Gemini. Given the small number of GCs, we build the GCLF by simply subtracting the background objects from the magnitude histograms. However, for the LSBds 13, 16 and 22 (which appear to have more GCs), we tentatively fit the GCLF, as shown in Fig. \ref{fig:13dist}. We derive an estimate for the number of GCs and specific frequency in each galaxy, obtaining the values N$_{GC}$ = 6.9, 8.7 and 17.57, S$_{N}$ = 45.0, 67.0 and 643.71, respectively, and the corresponding GCLF turnovers are 24.5$_{-3.1}^{+0.4}$, 24.2$_{-4.7}^{+0.6}$,  24.3$_{-0.5}^{0.4}$.

\begin{figure}
    \centering
    \includegraphics[width=1\linewidth]{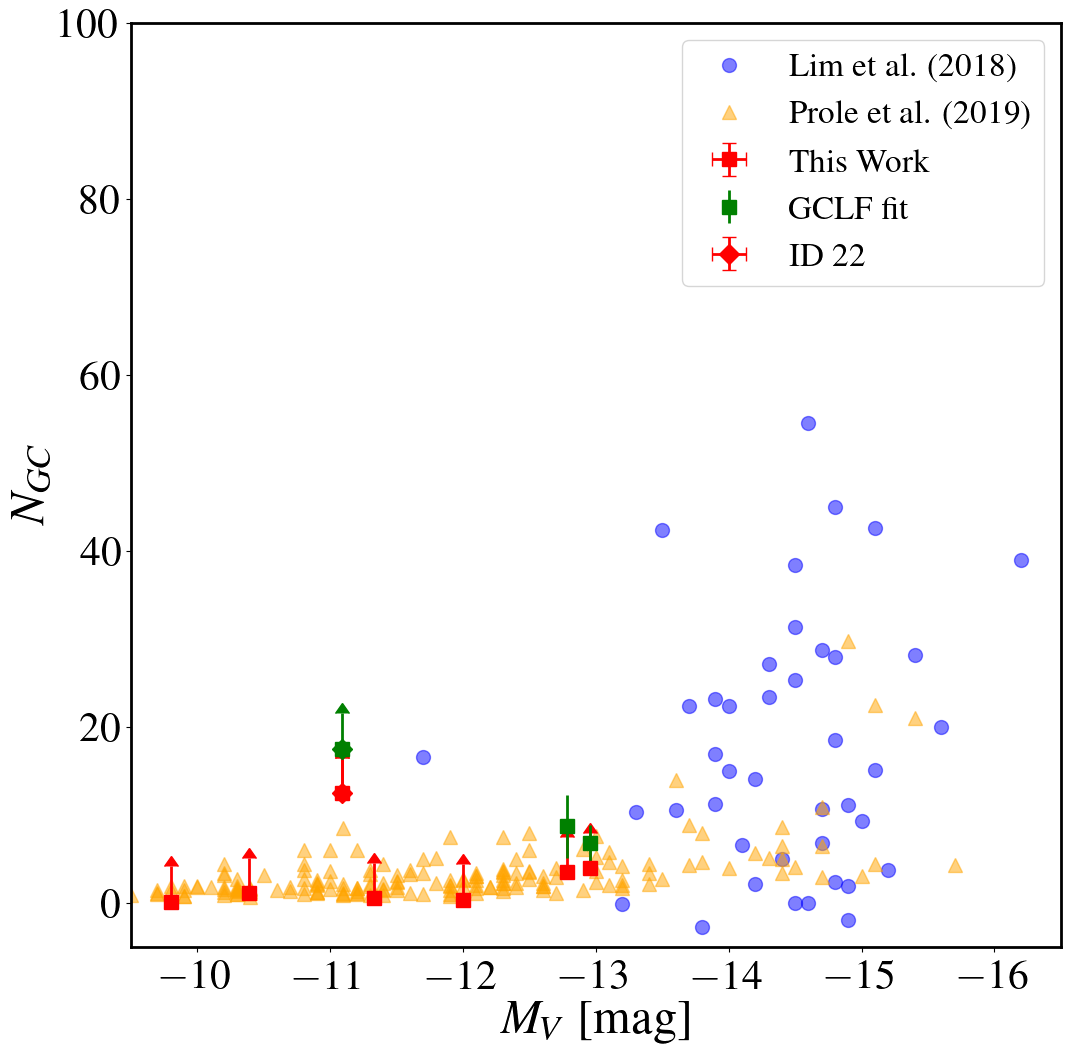}
   \caption{The lower limit for the $N_{GC}$ as a function of absolute V-band magnitude for the GC systems of the LSBds analysed around NGC 3115 (this work). For LSBds 13, 16 and 22 we also show the results from the GCLF fit (see text for details). Literature values for Coma cluster dwarfs \protect\citep{Lim2018}, Fornax cluster dwarfs \protect\citep{prole19a}, and LSB dwarf 6 from \protect\cite{sharina05} are also shown.}

    \label{fig:ngcs}
\end{figure}

In \autoref{fig:ngcs} we show $N_{GC}$ as a function of the absolute V-band magnitude $M_V$ for the respective host LSB dwarfs around NGC 3115 together with literature data.
With the exception of LSB 22, our dwarf candidates are consistent with the literature sample, especially that of \cite{prole19a}, following a distribution similar to a power-law.
We note however, that LSB dwarf 22 is close to the elliptical galaxy 
MCG-01-26-016 at 70\,Mpc \citep{costa98}, as can be seen in (\autoref{fig:gcs3}). Thus, the number count of GCs is overestimated due to the contamination from the GC system of the larger galaxy. 

Using the results of the GCLF fit, we calculate the specific frequency ($S_N$). The Specific Frequency was recovered using the following equation:
    \begin{equation}
       S_{N} =  \int^{\infty}_{-\infty} a\cdot e^{-\frac{(x-\mu)^2}{(2 \cdot \sigma^2)}} \cdot (10^{0.4\cdot (M_V+15)})dx.
    \end{equation}
    
\begin{figure}
    \centering
    \includegraphics[width=0.99\linewidth]{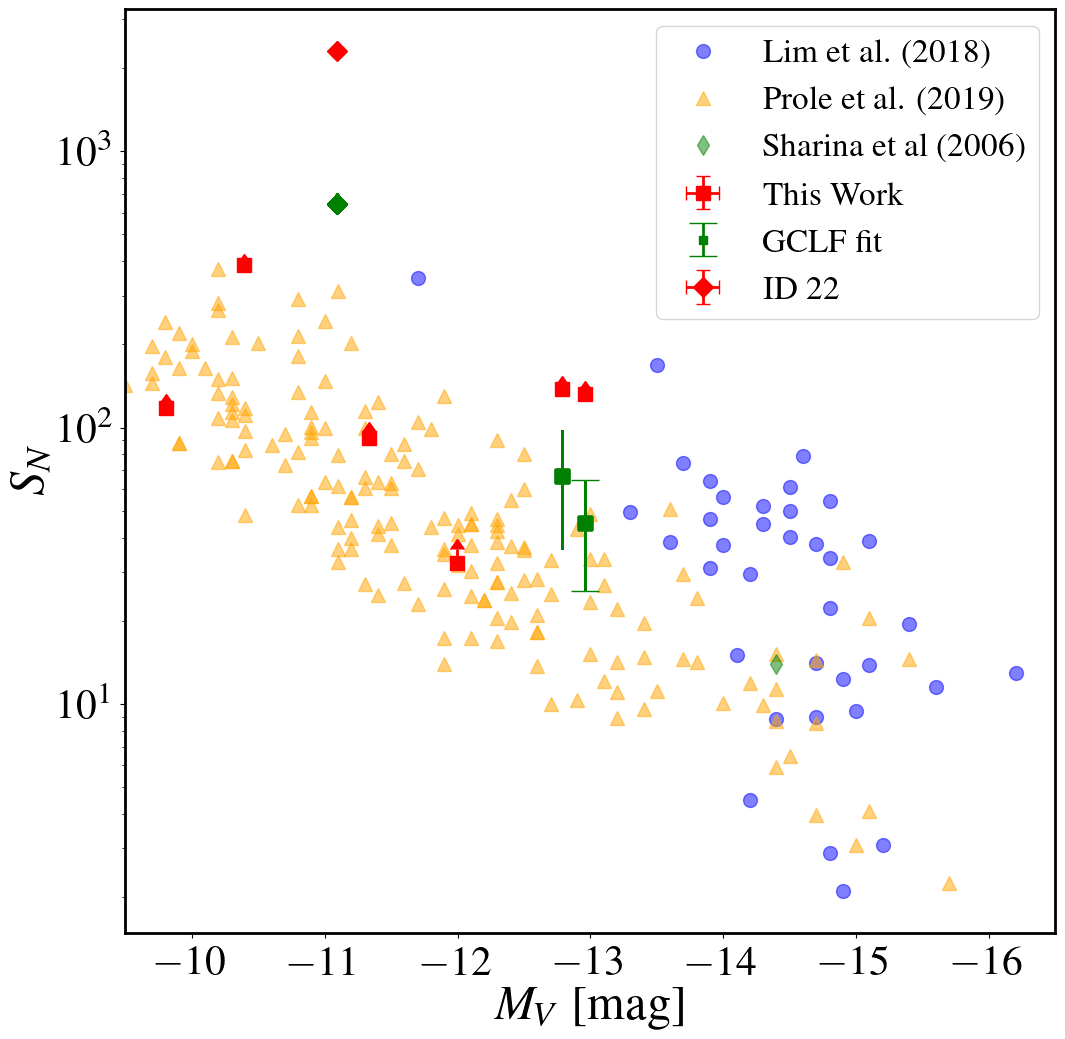}
    \caption{Similar to Fig.\ref{fig:ngcs} for the specific frequency (S$_{N}$) of the GC systems of the LSBds analysed around NGC 3115 (this work), the result from the GCLF fitting for LSBds 13, 16 and 22, Coma cluster dwarfs  \protect\citep{Lim2018}, and Fornax cluster dwarfs  \protect\citep{prole19a} and LSB dwarf 6 from  \protect\cite{sharina05}.}
    \label{fig:specific_frequency}
\end{figure}

In Fig. \ref{fig:specific_frequency}, we present the $S_N$ values for the GC systems of the LSBd with the ID 6 (based on Sharina et al. 2005), 13, 16, 17, 18, 19, 22, and 23. The LSB dwarfs ID14 and ID24 are compatible with not having GCs.

To facilitate comparisons, we converted the absolute $g$-band magnitude ($M_g$) to the absolute $V$-band magnitude ($M_V$) using Equation \ref{eq:lupton_V}. Notably, our sample of GC systems are found at a similar loci as the Fornax LSB dwarfs GC systems \citep{prole19a} of the same brightness ($M_V$).

There are two particular LSBds, ID 22 and ID 19, that appear to exhibit notably high $S_N$ values. It is important to clarify that ID 22 is affected by contamination from a nearby galaxy, while the high $S_N$ value in the case of ID 19 is due to the model not fitting well, primarily because of the limited number of GC candidates available for analysis in this particular LSBd. Although the N$_{GC}$ and $S_N$ are systematically higher than previous studies, within the uncertainties they are roughly consistent with the overall distribution of previous studies.

\section{Discussion}
\label{sec:discussion}

\subsection{Environment and the lack of high-mass LSB dwarfs}
The environment appears to exert a significant influence on the characteristics exhibited by the populations of LSBds and UDGs. A noteworthy pattern emerges when examining their colours: red LSB dwarfs are predominantly found in cluster environments, while blue counterparts tend to dominate in the field \citep{prole19a, Martin19, prole19}. Interestingly, we found that our population tends to display redder colours than expected of field LSBds, as shown in Fig. \ref{fig:colour_hist}. This might be due to the fact that NGC3115 is a rare isolated S0 galaxy and some mechanism may have acted to quench the galaxy and its halo environment (see below). 

While this occurrence may appear unexpected, other studies have also reported the presence of isolated red UDGs \citep{md2016,roman2019}. Moreover, the largest LSBd satellite of NGC3115, known as ID6, is one of the closest objects to the host and is also one of the reddest. It is worth noting that this particular object would not meet the UDG criteria, as its size measures approximately $\sim$0.6 kpc, which is at the lower limit of UDG sizes (typically $R_e \gtrsim 1.5$ kpc). 

Dwarf galaxies play a crucial role in understanding the processes that shape the evolution of galaxies. As such, they serve as valuable tools for studying and tracing the impact of these processes in cosmology. In a recent study by \cite{Watkins23}, they explore how dwarf galaxies can shed light on the ongoing debate regarding their influence on the relationship between galaxy size and mass. In our analysis, we observed that the LSBds projected closer to the center tend to exhibit a redder colours. This observation may suggest that these centrally located LSBds may be experiencing a quenching of star formation due to the depletion of their gas reservoirs in their local environment. 

In \citet{Watkins23} it was discovered that there are clear associations between the stellar mass of dwarf galaxies with their surface brightness and central stellar density. It was observed that lower-mass dwarf galaxies typically exhibit lower values of both surface brightness and central stellar density. This suggests a connection with the gravitational potential wells within these galaxies, which diminish as the stellar mass decreases. Additionally, it highlights the significant role of feedback mechanisms in the removal and redistribution of mass within low-mass dwarf galaxies as opposed to their high-mass counterparts. This phenomenon potentially leads to the formation of more diffuse structures in low-mass dwarf galaxies \citep[e.g][]{Governato10, dicintio17, Watkins23}. Furthermore, it may provide an explanation for the scarcity of high-mass low-surface-brightness dwarf galaxies, as these galaxies are more likely to accumulate central mass due to being less susceptible to the feedback processes described. In another study conducted by \citet{Junais22}, they investigated a sample of LSB galaxies within the Virgo cluster. The work shows that the majority of LSB galaxies appeared predominantly red, which is consistent with expectations in a cluster environment. Furthermore, they discovered evidence of a relationship between the colour of these galaxies and their distance from the cluster center. In particular, their observations revealed that LSBds, which exhibit a blue, star-forming appearance, are typically located at greater distances from the cluster center. This aligns with the results we have presented for the colour distribution of the LSBds in Fig. \ref{fig:colour_hist}. 

By analysing their HI gas content, \citet{Junais22} suggest that these galaxies may have undergone significant gas loss due to strong ram pressure stripping (RPS), with a \textit{ram pressure stripping}  time varying according with the cluster-centric distance, older RPS event are close to the centre. Based on \cite{Junais22}, we have the opportunity to establish a relation between their findings on the mechanisms of formation and the relation between colour and distances of these LSBds from the host galaxy with our LSBds. This analysis can provide valuable insights into their potential evolutionary pathways  a low-density environment. In such environments, RPS effects are less pronounced compared to those explored in the previously cited works. However, a recent study by \cite{Paudel23}, working with the S0 galaxy NGC 936 in a group environment, has detected compelling evidence of an ongoing disruption process caused by tidal forces exerted by the host S0 galaxy. This discovery lends support to the notion that tidal stripping mechanisms could indeed play a significant role in the formation of LSBds, as previously mentioned.

 \cite{Prole21a} examined the quiescent portion of isolated LSBds finding a distinct bi-modal population based on colour. The bluer LSBds tended to exhibit higher Sérsic indices and more concentrated profiles, with a peak at $n \sim 1$, a value commonly regarded as the standard for LSB dwarf galaxies in the existing literature. Conversely, the red population displayed a peak value at  $n \sim 0.7$, which closely aligned with values observed in UDGs  galaxy clusters. They also identified a correlation between colour and the environment, with blue LSBds being predominantly concentrated in low-density surroundings. Considering blue and red galaxies, our work, together with previous works in the literature, suggest that there seems to exist a marginal preference for bluer galaxies to have slightly larger Sérsic index than redder galaxies. Overall, our LSBds Sérsic indices are consistent with the ones found by \cite{Prole21a} with a peak at $n \sim 0.9$ for the bluer population and $n \sim 0.7$ for the red population. 
 
 The study of \cite{Prole21a} estimated that approximately $26\pm5\%$ of isolated local LSBds belonged to the red population, which is interpreted as the quiescent fraction. This finding challenged prevailing assumptions on the prevalence of quiescent LSBds, suggesting that while high-density environments may exert a dominant influence, they are not the exclusive factor in generating quiescent LSBds. Consequently, this could account for the presence of a redder population in the low-density environment surrounding NGC 3115.

Another crucial factor that can provide insights into the formation mechanisms of these dwarf galaxies is their kinematic properties. In the study conducted by \cite{CB2020}, they investigated isolated UDGs formed in the hydrodynamical simulation suite NIHAO and observed a diverse range of kinematic profiles, spanning from galaxies with dispersion-supported motion to those with rotation-supported motion. In \cite{Yaryura2023}, they found that groups associations of simulated LSB galaxies located in higher-density environments exhibited higher velocity dispersion compared to their counterparts in less dense environments. This result underscores the significant influence of the galactic environment on the dynamical properties that could affect the formation and evolution mechanisms of LSB galaxies. These studies emphasize the importance of acquiring dynamical information, and future Integral Field Unit (IFU) observations of our LSBds could offer valuable insights into whether they are pressure-supported or rotation-supported systems, thereby shedding light on their formation mechanisms and the influence of the environment on these mechanisms.

\subsection{GC Systems and Total Mass}

LSBds have been observed to host diverse populations of GCs, as shown by several studies \citep[e.g.][]{amorisco18, prole19a, Muller2021}. In the Coma cluster, massive LSBs are particularly notable for their abundant GCs \citep{saifollahi22}. In the Hydra cluster, UDGs generally exhibit compact sources, and the number of GCs varies  the range of 3 to 10 \citep{iodici2020}. Conversely,  the Perseus cluster, the GC populations among UDGs display diversity, with some being GC-rich and others GC-poor \citep{gannon22}. 

When examining UDGs  the Virgo cluster, their GC systems exhibit a wide range in specific frequency (S$_N$). On average, Virgo UDGs possess a higher $S_N$ than typical Virgo dwarf galaxies but a lower S$_N$ compared to Coma UDGs at equivalent luminosity levels \citep{lim2020}. It is noteworthy that while the GC systems in UDGs are primarily composed of blue clusters, the contribution of red clusters is more significant in the more massive UDGs.

In contrast, there is relatively limited research focused on low-density environments. A study of GC candidates  MATLAS UDGs situated in low to moderate-density environments found no compelling evidence of a higher GC specific frequency in UDGs compared to classical dwarf galaxies \citep{marleau21}. Additionally, a notable observation across various studies is the dispersion observed in the specific frequency of GC systems among both dwarf galaxies and UDGs, particularly at the faint end \citep{Lim2018, prole19a, Muller2021}.

The analysis of GCs also raises concerns about potential contamination in our LSBd sample, suggesting the possibility of some dwarfs being background or foreground galaxies. 
As observed in Fig. \ref{fig:observations}, objects 20, 22, 23, and 24 cluster in the north-central region of our observed field of view.  this region, three larger galaxies share similar redshifts (approximately $\sim 0.016$), identified as MCG -01-26-016, MCG -01-26-015, and MCG -01-26-017 (NED).

To determine whether these objects are part of the NGC\,3115 satellite system or belong to another system at a redshift of approximately z$\sim0.016$, follow-up observations are required. It is worth noting that LSBd 22 is nucleated and surrounded by GC candidates that, based on our analysis, may be overestimated due to contamination from the nearby massive galaxy. If we were to place LSBd 22 at z$\sim 0.016$, the size and absolute magnitude of such system would measure approximately $\sim$2.69 kpc and Mg $\sim$-14.66, respectively. If the distance of 64 Mpc is confirmed, LSBd 22 would be classified as a UDG.

\subsubsection{Total Mass Upper Limit Estimation}

Identifying the population of Globular Clusters (GCs)  a galaxy holds significance because it allows us to estimate the total mass of the host galaxy \citep{Beasley2016b,harris2017, forbes18, Forbes2020a, zaritsky22}. This estimate is based on the observed relationship between the number of GCs per galaxy (N$_{\rm GC}$) and the total galaxy mass (M$_{\rm T}$), as shown in previous studies \citep{forbes18, Forbes2020a, zaritsky22}. This relationship has been explored across various mass ranges and for different types of galaxies.

When specifically examining the mass of the GC system (M$_{\rm GC}$), previous research by \cite{harris2017} and \cite{forbes18} established relationships between M$_{\rm GC}$ and M$_{\rm T}$ for their respective host galaxies. However, \cite{Forbes2020a} further confirmed the utility of N$_{\rm GC}$ as a reliable tracer of the virial mass of the host galaxy's halo. Importantly, this relationship is linear and remains consistent across various galaxy morphologies, extending its applicability to dwarf galaxies regime. This is significant because N$_{\rm GC}$ provides a more reliable estimate correlating more strongly with M$_{\rm GC}$ \citep{Harris_2013, Forbes2020a, saifollahi22}.

Recent work conducted by \cite{zaritsky22} has unveiled a nearly linear correlation between the number of GCs per galaxy (N$_{\rm GC}$) and the total galaxy mass (M$_{\rm T}$) for low-luminosity galaxies. Specifically, they found that N$_{\rm GC}$ scales as N$_{\rm GC} \propto $ M$_{\rm T}^{0.92 \pm 0.08}$ for galaxies with total masses down to M$_{\rm T} \sim 10^{8.75}$ M$_{\odot}$.

\begin{figure}
    \centering
    \includegraphics[width=1\linewidth]{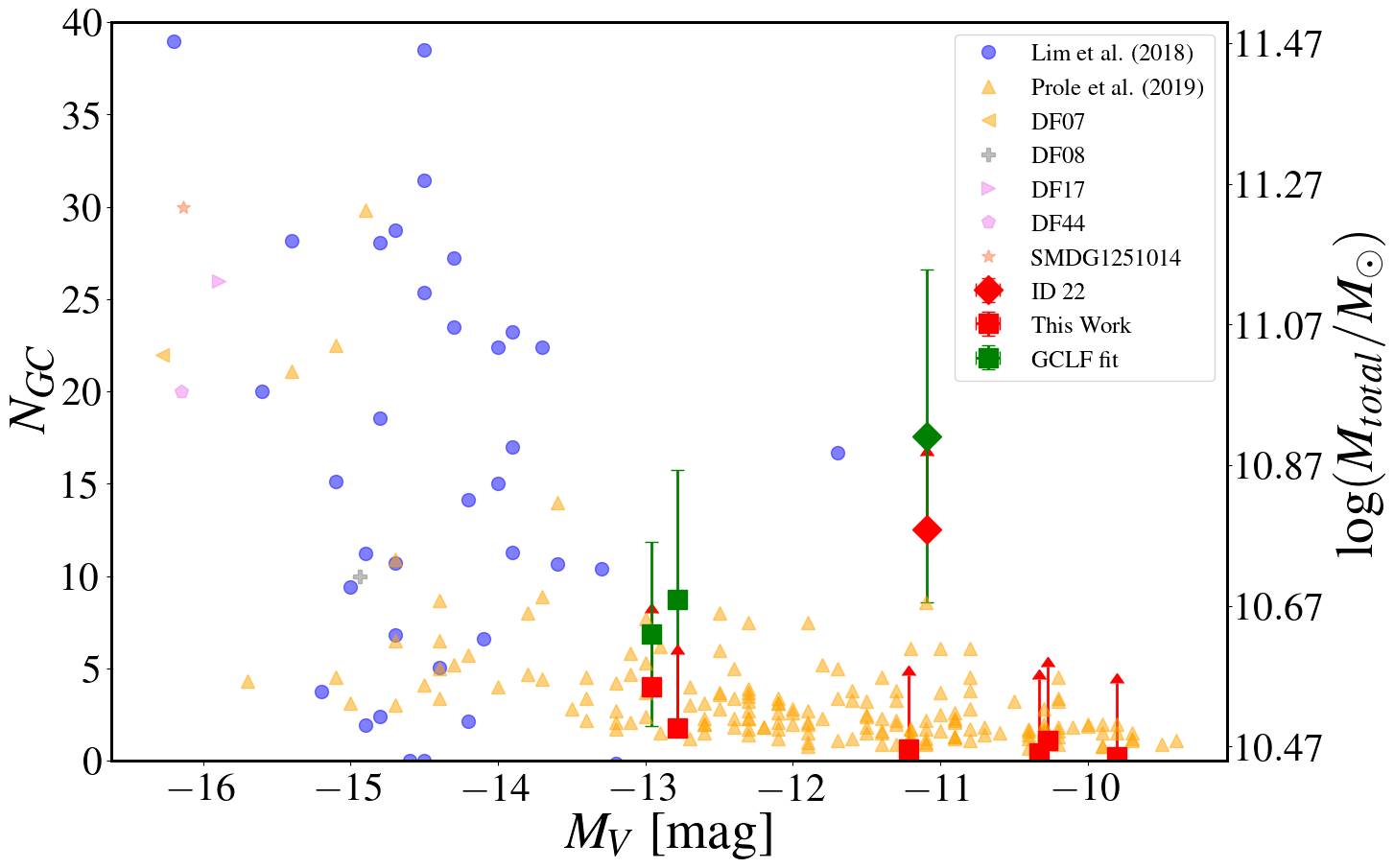}
        \caption{$N_{GC}$ as a function of the absolute magnitude in the V band (M$_{V}$) with the total mass upper limit (M$_{T}$) obtained for the LSBds identified around NGC 3115 (this work), Coma Cluster dwarfs \protect\citep{Lim2018}, Fornax Cluster dwarfs \protect\citep{prole19a}, LSB dwarf 6 from \protect\cite{sharina05} and the six UDGs identified by \protect\cite{saifollahi22} in the Coma Cluster.}
    \label{fig:total_mass}
\end{figure}

By applying the N$_{GC}-$M$_{\rm T}$ relationship established by \cite{zaritsky22}, we were able to estimate the upper limit of the total mass for the host LSBds, for which we had acquired their N$_{\rm GC}$ values (detailed in Sec. \ref{sec:GCs}). Notably, there is an outlier in our dataset, specifically LSBd 22, which can be attributed to an overestimation of the GC candidates in this galaxy. Excluding LSBd 22 from our analysis, we obtained the mean total mass value of approximately log(M$_{\rm T} $) $ \sim 10.5 M_{\odot}$, with the upper limit for the total mass ranging from $\sim 10.4$ to $\sim 10.5 $ M$_{\odot}$. In Fig. \ref{fig:total_mass}, we present the upper limit of the total mass for our LSBd sample and compare it with results obtained for Coma Cluster dwarfs \citep{Lim2018}, Fornax Cluster dwarfs \citep{prole19a}, LSBd 6 from \cite{sharina05}, and UDGs from the Coma Cluster identified by \cite{saifollahi22}. Remarkably, some objects in our LSBd sample present values  the range found for the Fornax cluster dwarfs, which are situated 20 Mpc away from the Milky Way. 

Our investigation has revealed that the N$_{\rm GC}$, S$_{\rm N}$, and the upper limits of total mass for our LSBds align with findings from other dwarf galaxies, showing a strong concurrence, particularly with the Fornax Cluster dwarfs. Additionally, the reasonable model obtained for the Globular Cluster Luminosity Function (GCLF) corresponds well with those found in dwarf galaxies.

\section{Summary and Conclusions}
\label{sec:summary}
In this study, we conducted a search for LSBds around the low-density environment surrounding the nearby S0 galaxy NGC\,3115 and their GC systems. We utilized deep $g$ and $r$ imaging from the Dark Energy Camera (DECam) and employed the DECam pipeline for data reduction. Subsequently, we conducted the photometry using \texttt{IMFIT} and \texttt{SExtractor} on objects classified as candidates for Low Surface Brightness dwarf galaxies through visual inspection. Leading to the identification of a final catalog comprising 24 LSBd candidates, in which 18 of them are reported for the first time. Through a comparative analysis with previously identified LSBds and those discovered during the course of our research, we have observed consistent findings and outcomes.

Furthermore, we followed-up observations of nine LSBds using GMOS deep g, z and i imaging that were reduced using Gemini DRAGONS, which enabled us to discern the Globular Cluster (GC) systems associated with these LSBds and gather significant insights into their host galaxies. When we compared our findings for the NGC 3115 system with those reported in previous studies, we observed that our sample aligns similarly with the characteristics of dwarfs in the Fornax Cluster. The primary outcomes of our investigation are summarized as follows:
\begin{enumerate}

\item  In the size-radius diagram (Fig. \ref{fig:money_plot}), the LSBds studied here fall in the same loci as the faint end of \cite{carlsten2020a} and \cite{eigenthaler18} sample and do not reach the loci of the Coma cluster UDG sample from \citet{vD15}.

\item As depicted in Fig. \ref{fig:money_plot}, most candidates display red colours. However, they are slightly smaller and fainter than what has been reported as the UDG limit (R$_{e} > $1.5Kpc, M$_g \sim-15$). 
 \item We uncover 24 LSB dwarf galaxies through visual inspection and photometry, of which 6 have been previously detected. 
\item Fig. \ref{fig:colour_hist} shows that our sample comprises LSBds that exhibit a range of colors. ($-0.2<g-r<1$), which is compatible with previous studies of LSB dwarf galaxies following the colour range found in \cite{carlsten2020b, Tanoglidis20} and \citep{Prole21a}. 
\item The luminosity function of the NGC\,3115 satellite system (Fig. \ref{fig:luminosity_function}) is comparable to that of the systems of NGC4258 and NGC4565
\item Our GC systems fall around the same loci of Fornax LSB dwarf GC systems \citep{prole19a} of the same brightness (M$_V$).
\item By applying the N$_{\rm GC}-$M$_{\rm T}$ relation by \cite{zaritsky22} we recover an upper limit for the total mass of these LSBds (Fig. \ref{fig:total_mass}), with the mean value of log(M$_{\rm T}$) $ \sim 10.5 $ M$_{\odot}$. 
\end{enumerate}
Despite LSBds being among the most numerous galaxies in the Universe, our understanding of them remains limited. Their formation and evolution are significantly influenced by the environment in which they reside in. It is worth noting that LSBds are frequently identified in high-density environments, primarily because their distances can be more easily determined in such regions, and also due to the challenge of detecting and measuring these faint systems. Estimating distances in the field, on the other hand, is more challenging and often requires extended exposure times. This factor is particularly crucial to consider as it introduces a bias into the study of such systems and can affect our comprehension of their formation and evolution. A comprehensive understanding of these pathways requires untangling the impact of the environment, which can be achieved by examining LSBds in low-density environments. In our study, we associated the distances of our LSBd sample with NGC 3115, a galaxy located in a low-density environment, enabling us to identify LSBds candidates within this specific context.

To understand the formation and evolution of the few known low surface brightness galaxies found in low-density environments we need a comprehensive study of their GCs populations \citep{vanDokkum19}. In summary, our findings, derived from the analysis of the GC population within these LSBds, bear similarities to the results obtained by \cite{prole19a}. They had previously identified LSBds within the Fornax Cluster Environment and observed a similar bimodal population of LSBds in terms of colour. However, a significant difference in our sample is that our LSBds tend to be redder compared to the typical expectation for LSBds in the field. Building upon the work of \cite{Paudel23} and the relationship we uncovered between colour and galactocentric distances of the LSBds relative to NGC 3115, our findings point towards a formation pathway influenced by tidal stripping mechanisms.

To further unravel the assembly processes and star formation timescales of LSBds within low-density environments, particularly through their GC systems, several critical steps lie ahead. Firstly, it is imperative to confirm the membership of the GCs within these galaxies, followed by inferring their stellar population parameters. Achieving this requires the acquisition of spectroscopic data.
Subsequently, by determining the mass, age and metallicity of their stellar populations, we gain a valuable opportunity to delve into the formation and evolution of these LSBds. These discoveries will allow us to refine our understanding of LSBd galaxy formation within low-density environments and establish a connection between LSBds and UDGs. Furthermore, with a more robust estimate of the total number of GCs in each galaxy and the direct measurement of velocity dispersion through spectroscopy, we can endeavor to establish an upper limit for the average ratio of halo mass to stellar mass (M$_{halo}/$M$_{stars}$). This estimation will provide insights into the amount of dark matter mass associated with each galaxy. Consequently, these findings will enhance our comprehension of LSB dwarf galaxy formation within low-density environments and may elucidate the relationship between LSB dwarfs and UDGs within this specific environmental context.

\section*{Acknowledgements}
We thank the referee for the relevant comments and suggestions that improved this work.
MCG acknowledges PIBIC/CNPq and Propesq/UFRGS.
ACS acknowledges funding from the brazilian agencies \textit {Conselho Nacional de Desenvolvimento Cient\'ifico e Tecnol\'ogico} (CNPq) and the Rio Grande do Sul Research Foundation (FAPERGS) through grants CNPq-403580/2016-1, CNPq-11153/2018-6, PqG/FAPERGS-17/2551-0001, FAPERGS/CAPES 19/2551-0000696-9 and L'Or\'eal UNESCO ABC \emph{Para Mulheres na Ci\^encia}.
CF acknowledges the financial support from CNPQ (processes 433615/2018-4 and 314672/2020-6) and FAPERGS (21/2551-0002025-3 ).
RF acknowledges Propesq/UFRGS.
Emilio acknoweledges CAPES, CNPq and FAPESP.
We thank Chris Conselice for useful discussions. 
Based in part on data collected at the Blanco Telescope, DECam via the time exchange program between The Southern Astrophysical Research (SOAR) and Blanco NSF’s NOIRLab. The SOAR telescope is a joint project of the Minist\'{e}rio da Ci\^{e}ncia, Tecnologia e Inova\c{c}\~{o}es (MCTI/LNA) do Brasil, the US National Science Foundation’s NOIRLab, the University of North Carolina at Chapel Hill (UNC), and Michigan State University (MSU).
Based in part on observations obtained at the international Gemini Observatory, a program of NSF’s NOIRLab, which is managed by the Association of Universities for Research in Astronomy (AURA) under a cooperative agreement with the National Science Foundation on behalf of the Gemini Observatory partnership: the National Science Foundation (United States), National Research Council (Canada), Agencia Nacional de Investigaci\'{o}n y Desarrollo (Chile), Ministerio de Ciencia, Tecnolog\'{i}a e Innovaci\'{o}n (Argentina), Minist\'{e}rio da Ci\^{e}ncia, Tecnologia, Inova\c{c}\~{o}es e Comunica\c{c}\~{o}es (Brazil), and Korea Astronomy and Space Science Institute (Republic of Korea). The data was processed using the DRAGONS (Data Reduction for Astronomy from Gemini Observatory North and South.
This research has made use of the NASA/IPAC Extragalactic Database, which is funded by the National Aeronautics and Space Administration and operated by the California Institute of Technology.

%%%%%%%%%%%%%%%%%%%%%%%%%%%%%%%%%%%%%%%%%%%%%%%%%%
%%%%%%%%%%%%%%%%%%%%%%%%%%%%%%%%%%%%%%%%%%%%%%%%%%
\section*{Data Availability}

The data underlying this article are available from the corresponding author, upon reasonable request. 
The raw data are available in the NOAO Science Archive (\url{http://archive1.dm.noao.edu/search/query/}) under program number 2017A-0911 for the DECam data; and in the Gemini Observatory Archive (\url{https://archive.gemini.edu/searchform} under program number GN-2020B-Q-223 and GS-2020B-Q-237 for the Gemini data.



\bibliographystyle{mnras}
\bibliography{refs} % if your bibtex file is called refs.bib

%\begin{thebibliography}{99}
%\end{thebibliography}

%%%%%%%%%%%%%%%%%%%%%%%%%%%%%%%%%%%%%%%%%%%%%%%%%%

%%%%%%%%%%%%%%%%% APPENDICES %%%%%%%%%%%%%%%%%%%%%

\appendix

\section{Reliability of recovered structural parameters from simulations and uncertainties estimation} 
\label{sec:mock}
\begin{figure}
    \centering
    \includegraphics[width=\columnwidth]{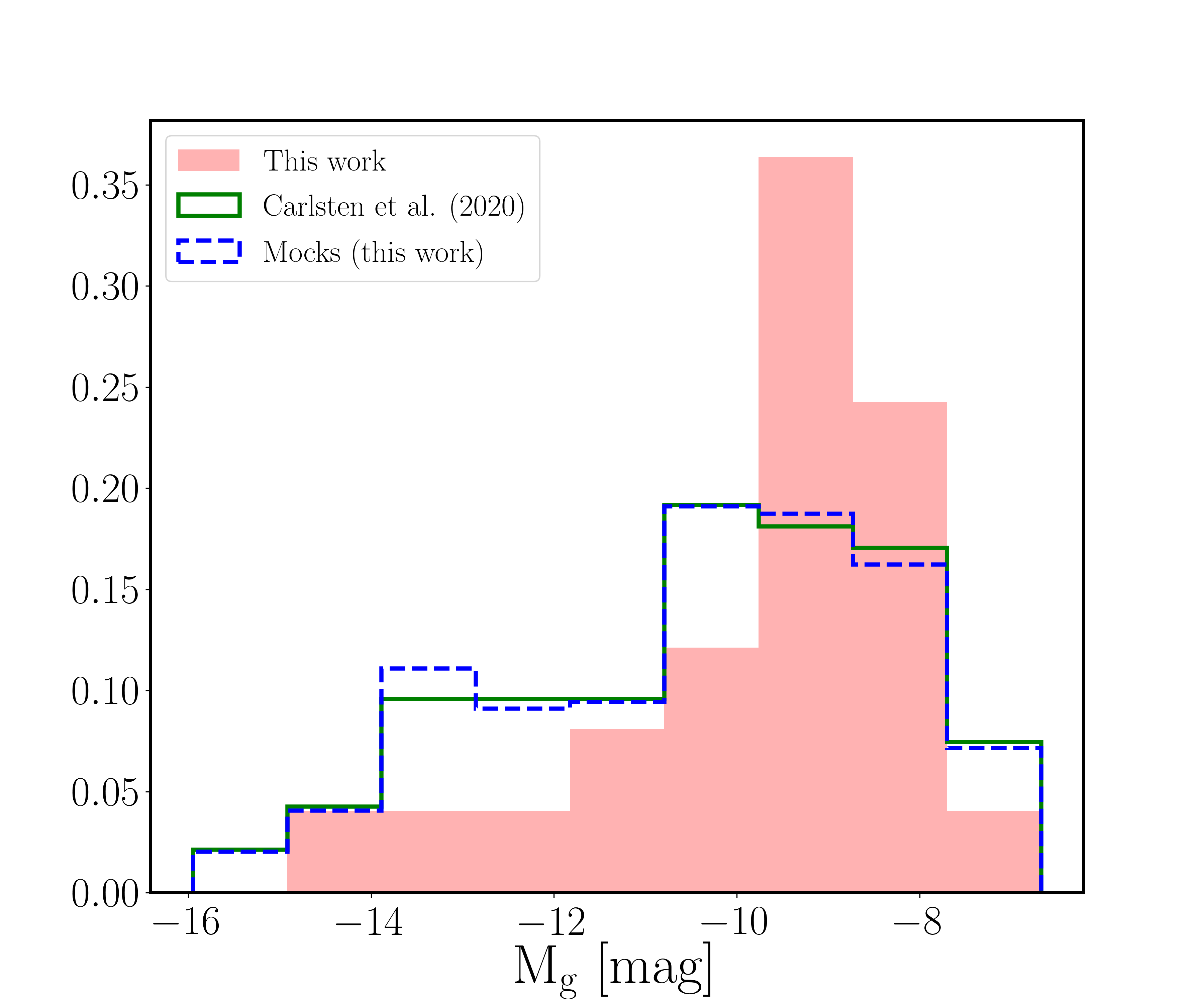} 
    \caption{Absolute magnitude density histograms for the LSB dwarf candidates of this work (red), the \protect\cite{carlsten2020b} sample restricted to $M_g \geq -16$ and $\rm R_e \leq 2.5 kpc$ (orange) and the mock galaxies (blue).}
    \label{fig:mag_dists}
\end{figure}

In order to address the reliability of photometric and structural parameters derived from the profile fitting described in Sec. \ref{sec:profile_fitting}, we generate a set of mock galaxies with known parameters assuming that their light distribution follows a single Sérsic model, then we try to recover them with a similar method. The mock galaxies are injected into randomly selected positions in the stacked images of the NE and SW pointings around NGC 3115. In order to generate a sample of mock galaxies with a realistic magnitude distribution, we recover the cumulative distribution function of $g$-band absolute magnitude from \cite{carlsten2020b} using the \texttt{ECDF} class of the \texttt{statsmodels}\footnote{\url{https://www.statsmodels.org/stable/}} Python library. Then we use the inverse transform sampling method to randomly sample the reconstructed empirical magnitude distribution for $\rm M_g \geq -16$. In Fig. \ref{fig:mag_dists} we compare absolute magnitude histograms of dwarfs from \cite{carlsten2020b}, the mock galaxies and the LSB dwarf candidates of this work. We choose \cite{carlsten2020b} as a reference for the magnitude distribution due to its moderate-sized sample of dwarf satellites in the Local Volume. For the $r$-band we assume the same magnitude distribution.
 
\begin{figure}
    \centering
    \includegraphics[width=\columnwidth]{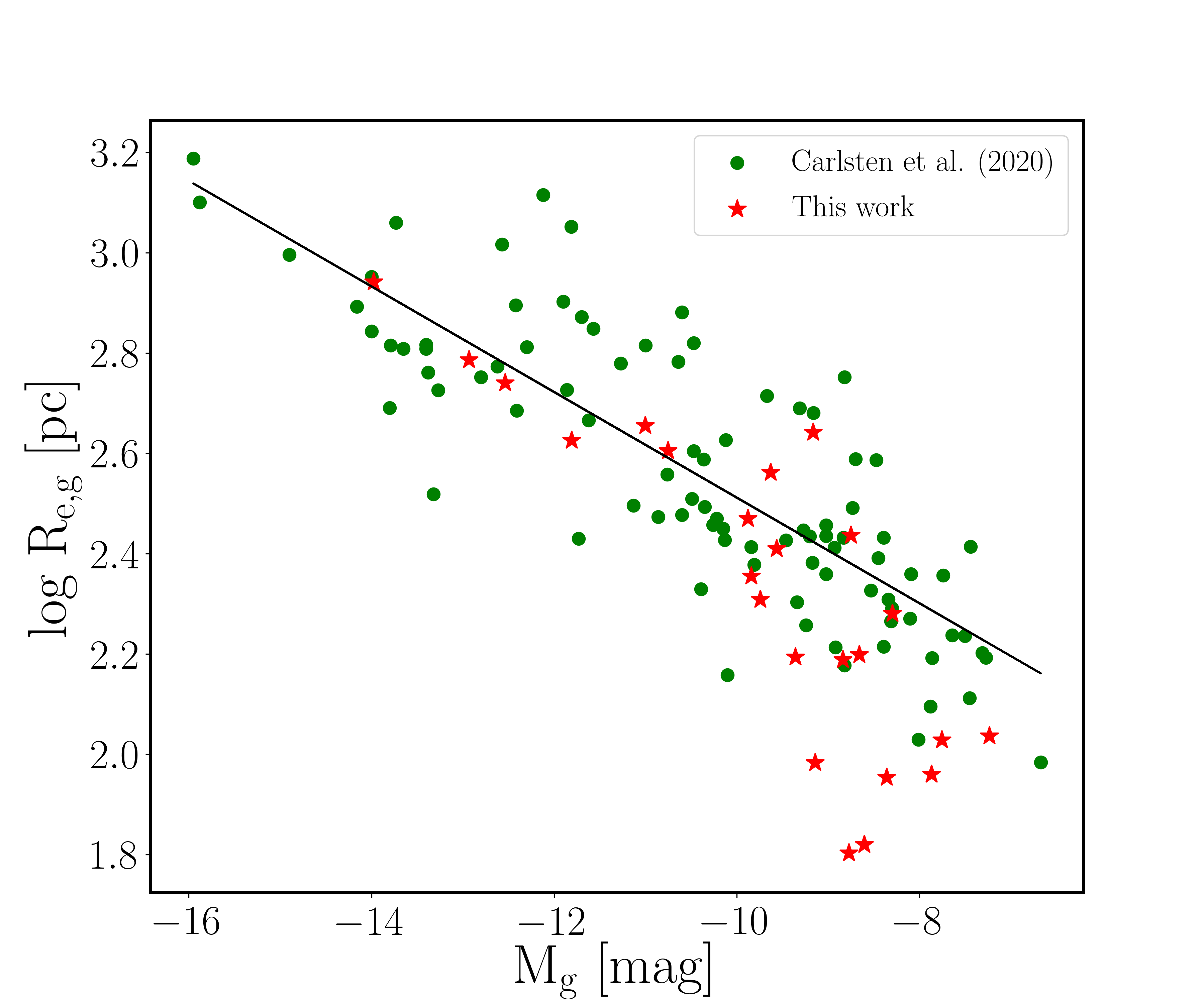} 
    \caption{Size-magnitude diagram showing the sample of this work (red) and the \protect\cite{carlsten2020b} sample restricted to $M_g \geq -16$ and $\rm R_e \leq 2.5 kpc$ (orange). The black line is the linear regression applied to \protect\cite{carlsten2020b} data, with coefficient $a = -0.105$ and intercept $b = 1.46$.}
    \label{fig:fit_re_M}
\end{figure}

\begin{figure*}
    \centering
    \includegraphics[width=0.19\linewidth]{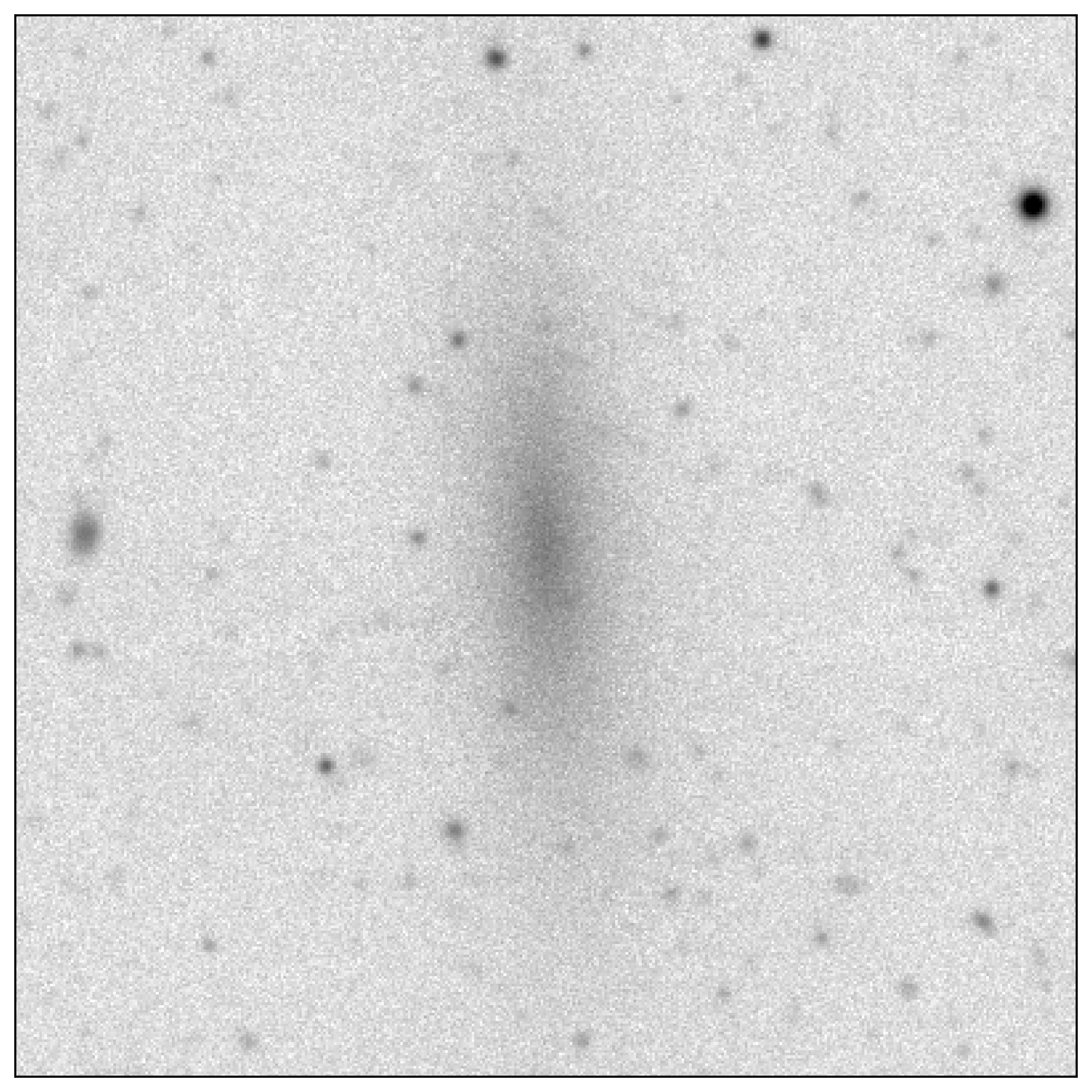} 
    \includegraphics[width=0.19\linewidth]{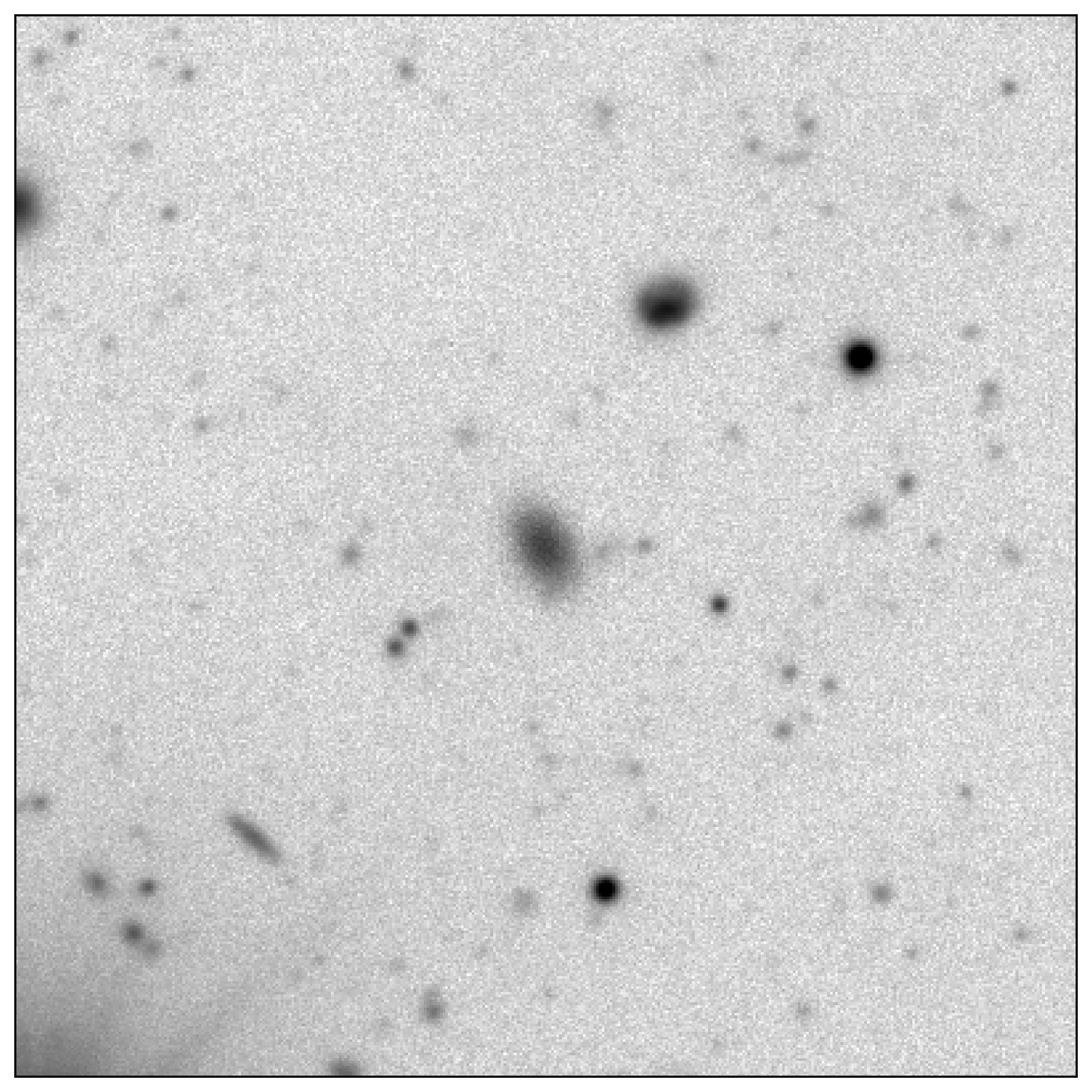} 
    \includegraphics[width=0.19\linewidth]{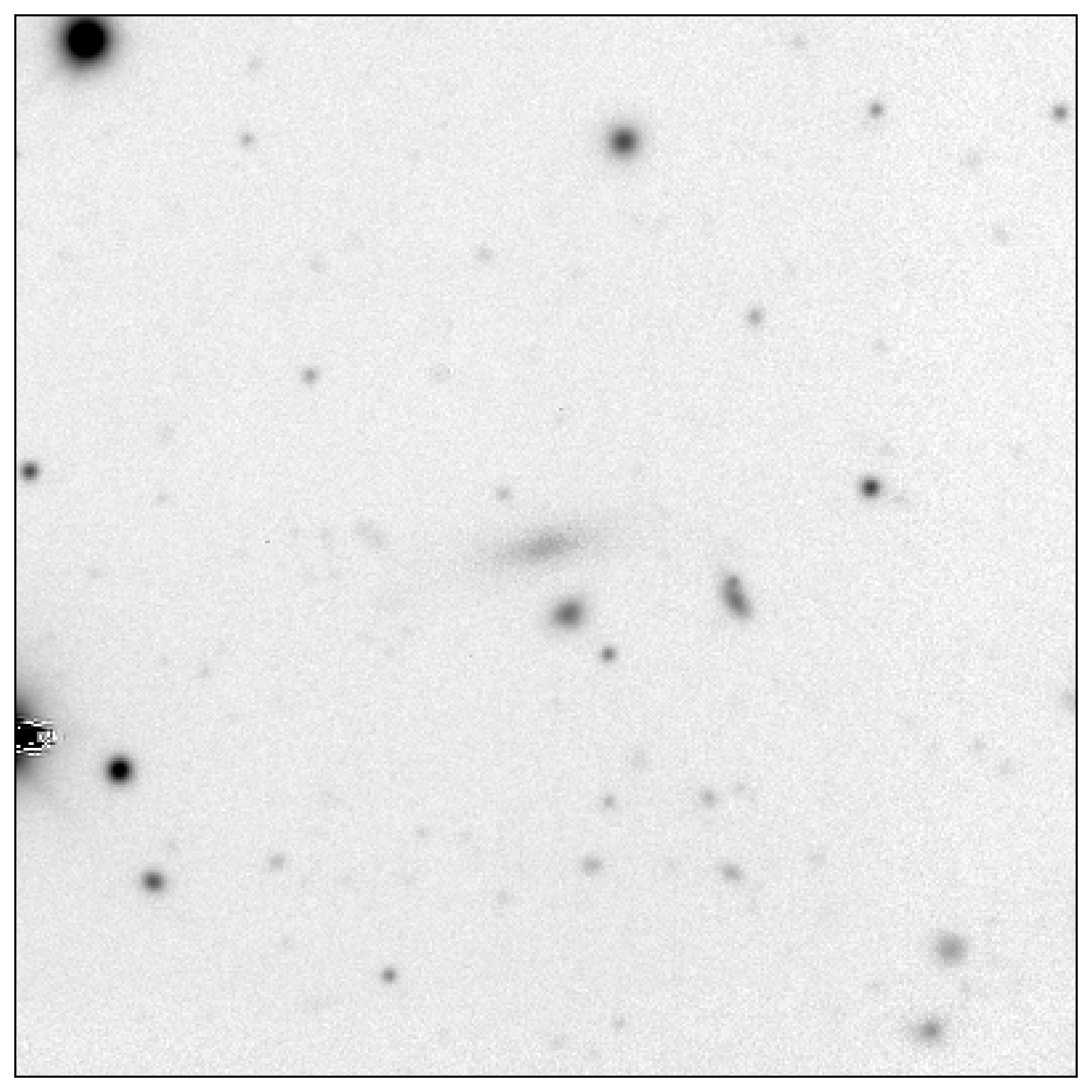} 
    \includegraphics[width=0.19\linewidth]{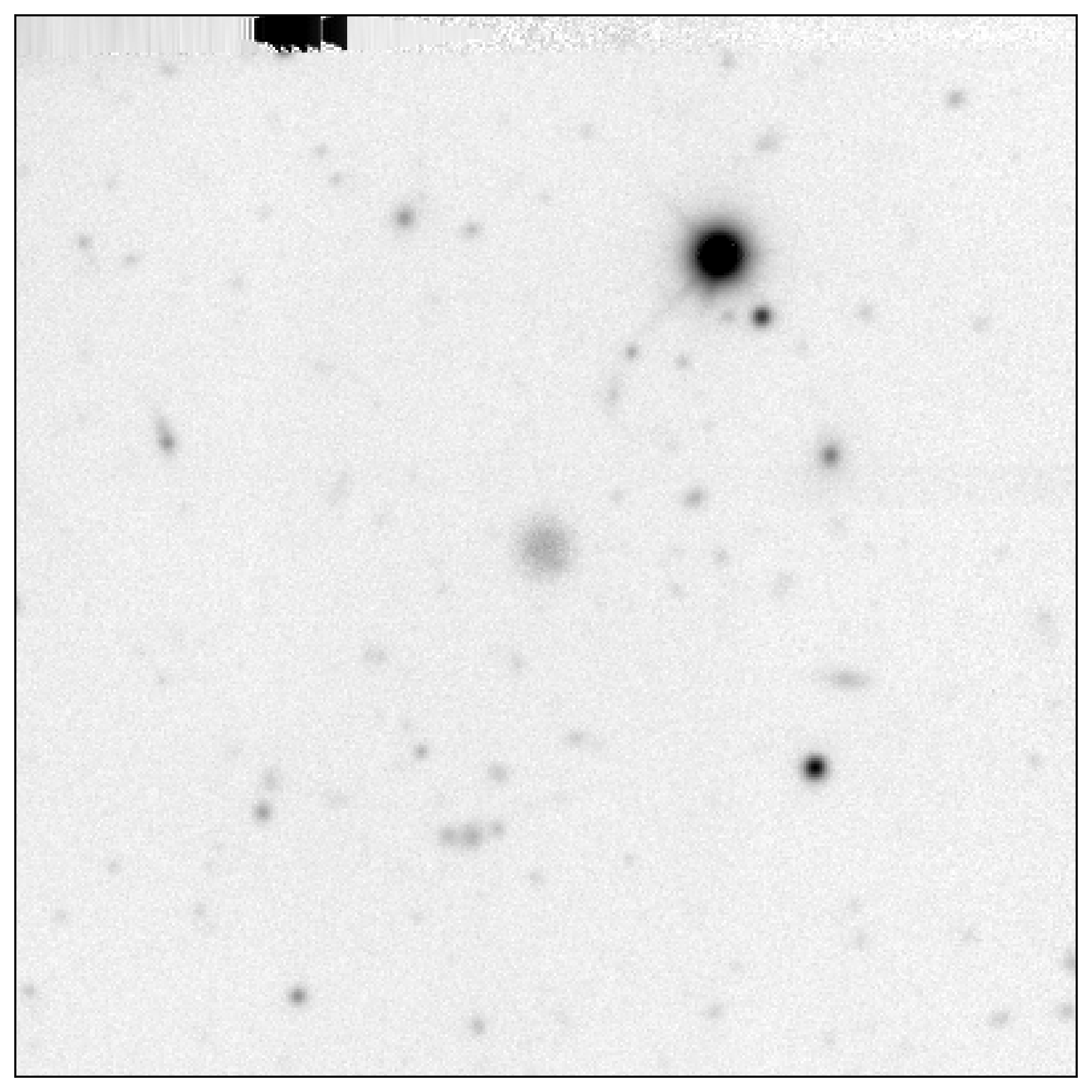} 
    \includegraphics[width=0.19\linewidth]{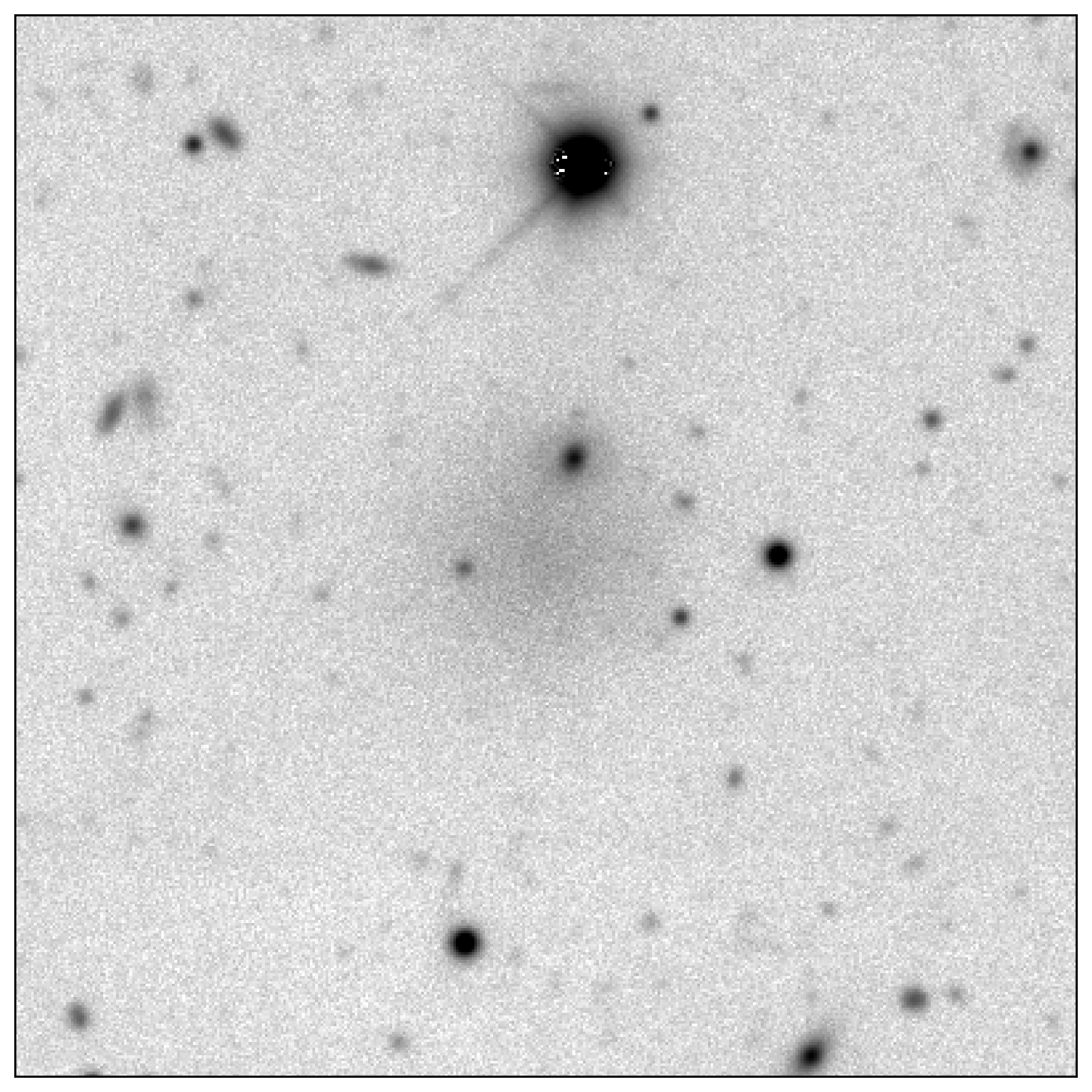} 
    \caption{Examples of mock galaxies. Images are in asinh scale and have the same dimensions of the stamps used during the fitting process (108 x 108 \arcsec).}
    \label{fig:mocks}
\end{figure*}

The effective radii of the mock galaxies are determined using the result of a linear regression applied over the \cite{carlsten2020b} $g$-band data (see Fig. \ref{fig:fit_re_M}). We assume the same relation for $\rm M$ and $\rm R_e$ of the $r$-band. We sample the absolute magnitude first, then compute the corresponding $\rm R_e$. We add a normal distributed noise in the computed $\rm R_e$, with standard deviation of $\Delta_{\rm max}$/2, where $\Delta_{\rm max}$ is the maximum difference between radii computed with the regression and the data from \cite{carlsten2020b}. It is important to note that all mock galaxies are simulated at the same distance as NGC 3115 (distance modulus of 29.93 mag and a scale of 46.94 kpc/$\arcsec$). The Sérsic index, PA and $\epsilon$ are sampled from uniform distributions with values ranging from 0.4 to 2, 0 to 180 degrees and 0 to 0.75, respectively. 
Once we have all the parameters, we generate their two-dimensional profiles with {\sc makeimage}, a companion program of {\sc imfit}. Then we convolve the models with the appropriate PSF, add Poisson noise and inject them into the stacked images. In Fig. \ref{fig:mocks} we show examples of mock galaxies generated. In total, 20000 mock galaxies are generated for each filter and their photometric and structural parameters are obtained according to the method described in Sec. \ref{sec:profile_fitting}, except for the iterative procedure to improve the fittings (steps 2 and 3).


\begin{figure*}
    \centering
    \includegraphics[width=0.49\linewidth]{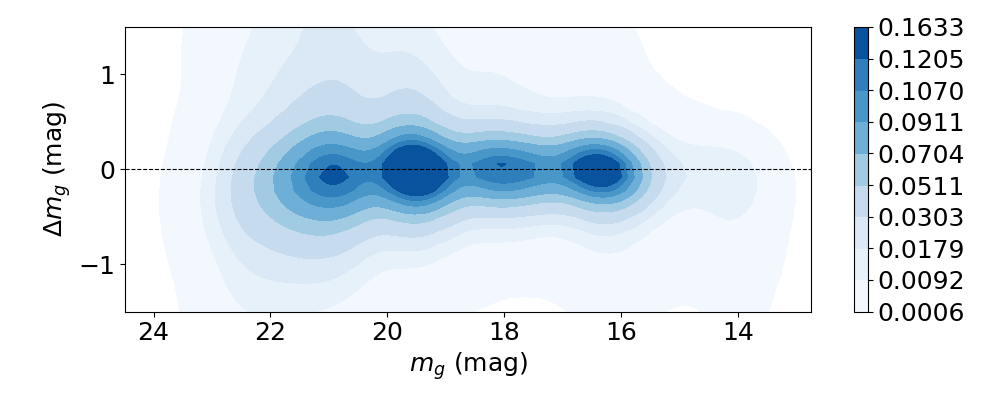}
    \includegraphics[width=0.49\linewidth]{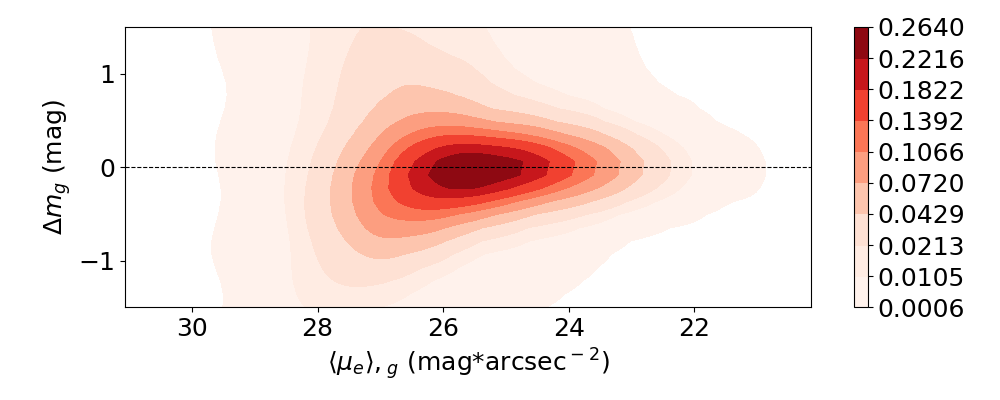}
    \includegraphics[width=0.49\linewidth]{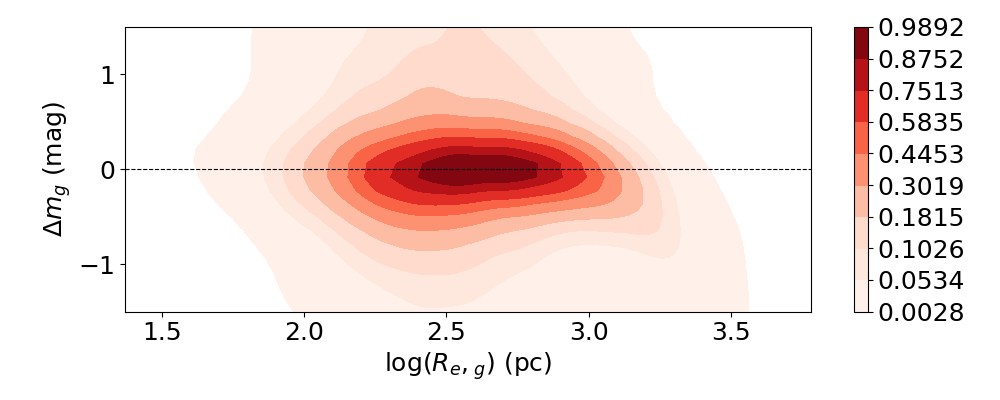} 
    \includegraphics[width=0.49\linewidth]{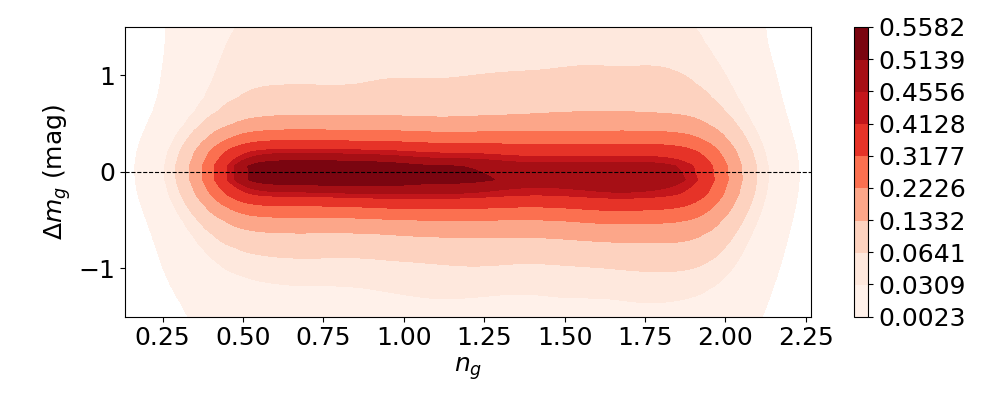}
    \caption{Kernel density estimation plots of the errors $\rm \Delta m_g = m_{g, true} - m_{g, fitted}$ versus all the parameters. The parameters in the horizontal axis are the true values.}
    \label{fig:kde_m}
\end{figure*}

\begin{figure*}
    \centering
    \includegraphics[width=0.49\linewidth]{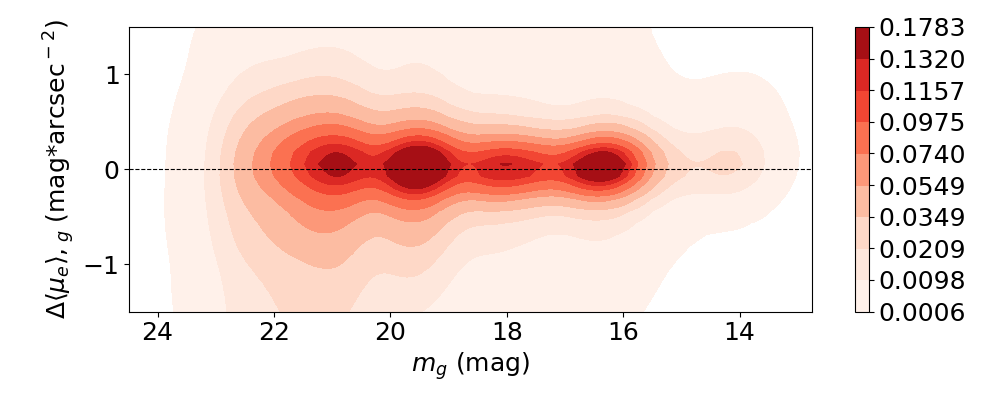}
    \includegraphics[width=0.49\linewidth]{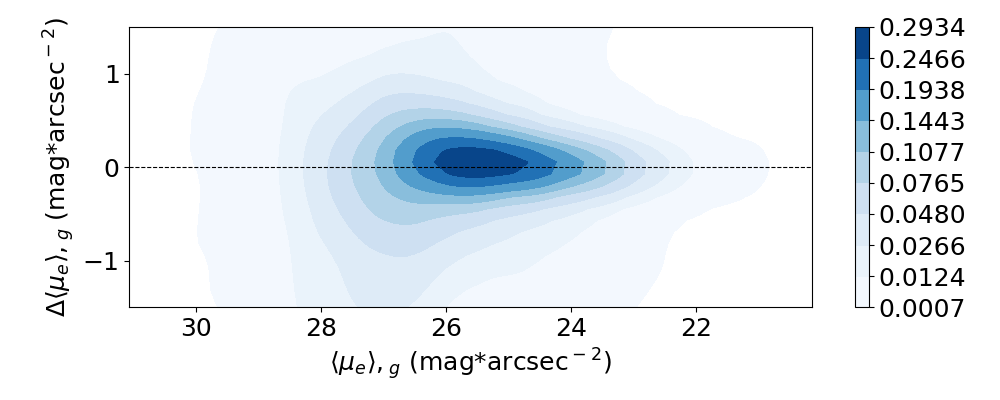}
    \includegraphics[width=0.49\linewidth]{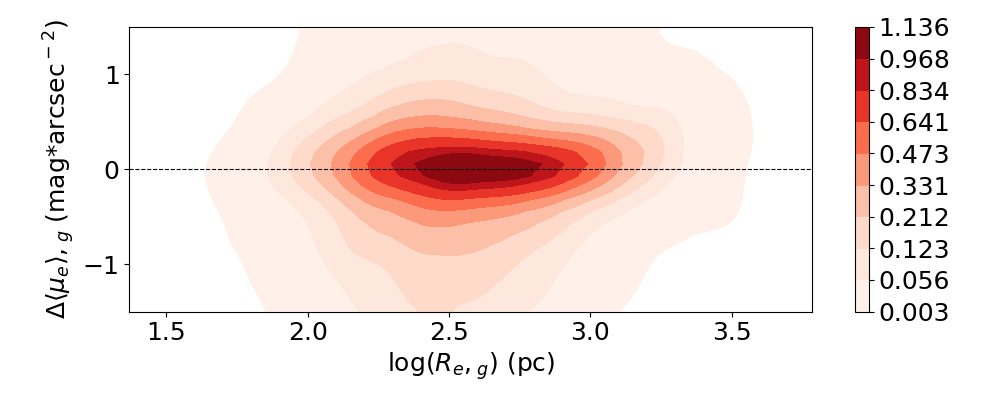} 
    \includegraphics[width=0.49\linewidth]{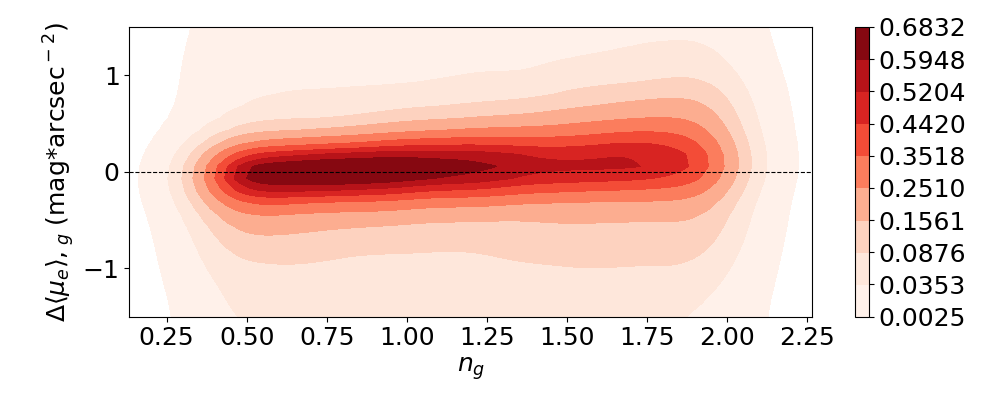}
    \caption{Same as in Fig. \ref{fig:kde_m}, but for surface brightness errors $\rm \Delta \langle \mu \rangle_{e,g}$.}
    \label{fig:kde_mu}
\end{figure*}

\begin{figure*}
    \centering
    \includegraphics[width=0.49\linewidth]{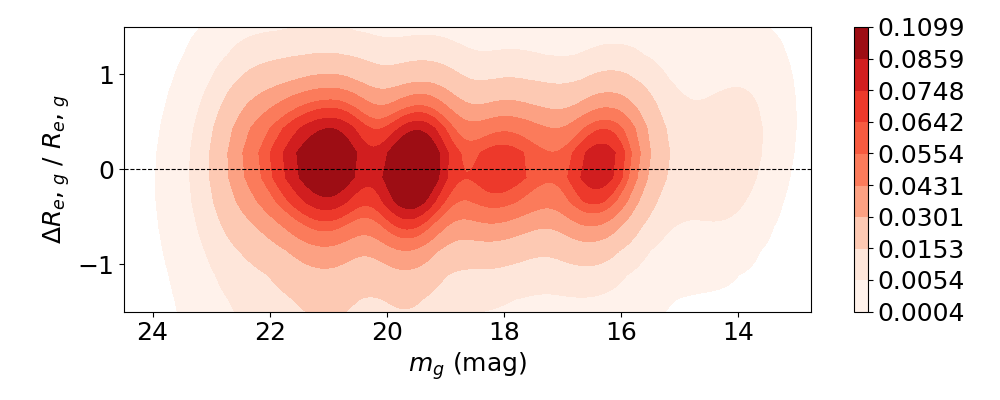}
    \includegraphics[width=0.49\linewidth]{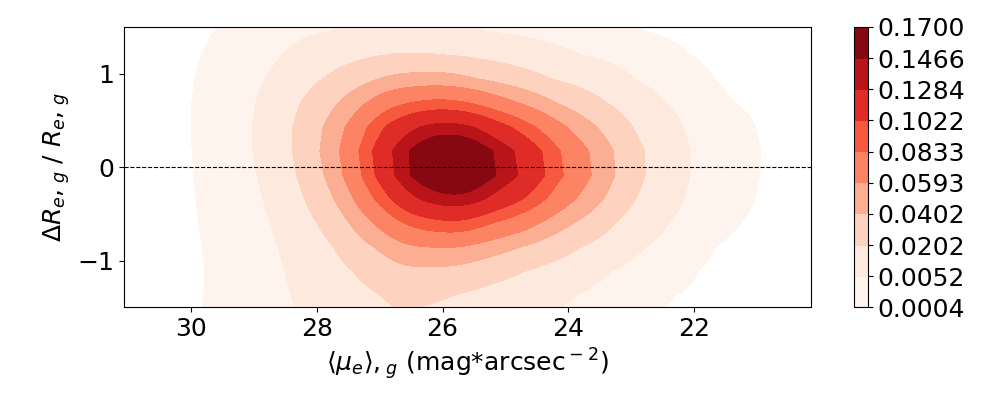}
    \includegraphics[width=0.49\linewidth]{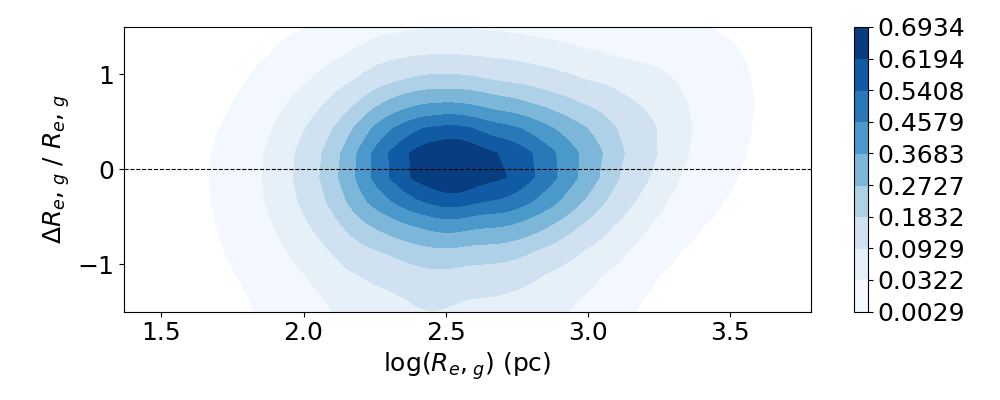} 
    \includegraphics[width=0.49\linewidth]{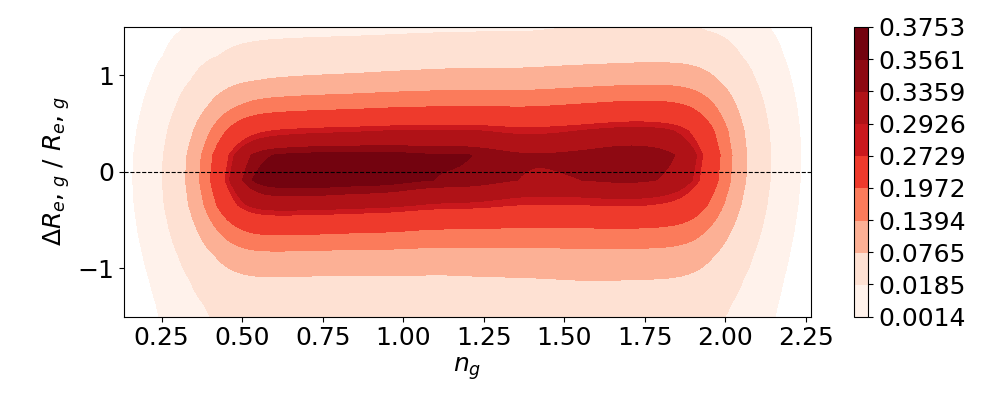}
    \caption{Same as in Fig. \ref{fig:kde_m}, but for effective radius percentage errors $\rm \Delta R_{e,g}/R_{e,g}$.}
    \label{fig:kde_re}
\end{figure*}

\begin{figure*}
    \centering
    \includegraphics[width=0.49\linewidth]{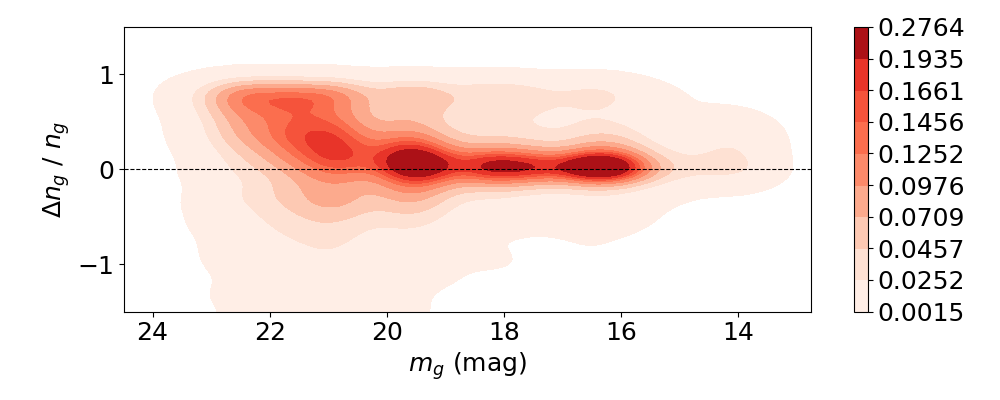}
    \includegraphics[width=0.49\linewidth]{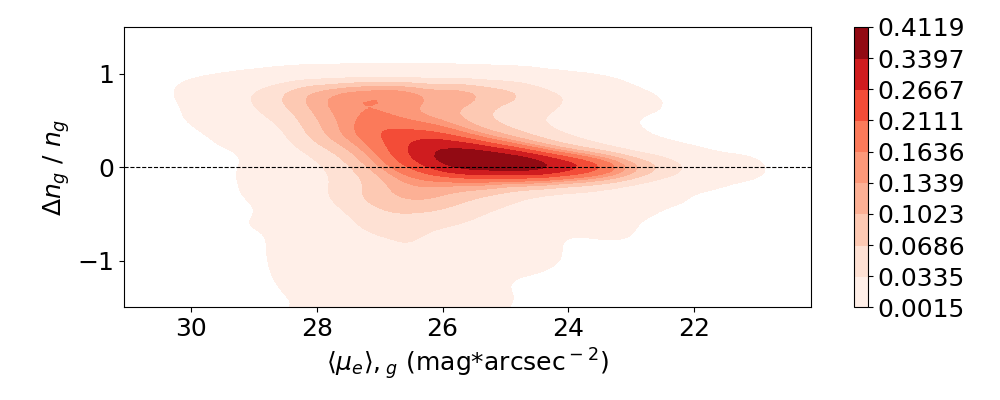}
    \includegraphics[width=0.49\linewidth]{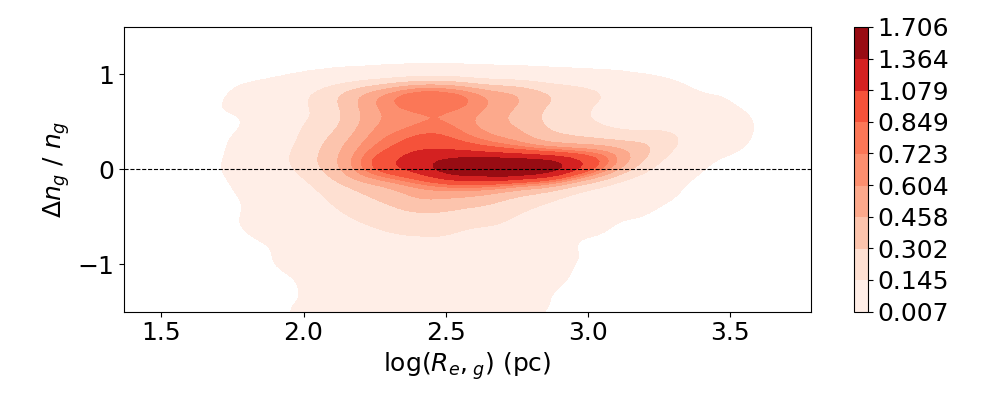} 
    \includegraphics[width=0.49\linewidth]{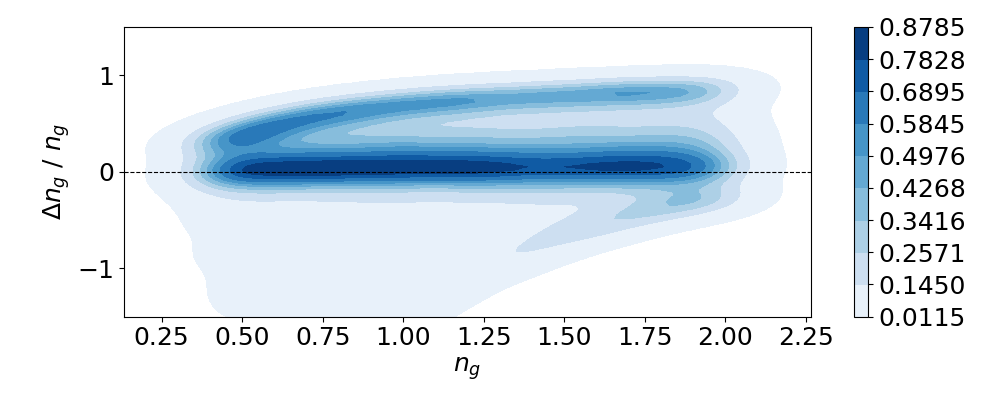}
    \caption{Same as in Fig. \ref{fig:kde_m}, but for Sérsic index percentage errors $\Delta n_g/n_g$.}
    \label{fig:kde_n}
\end{figure*}

We also use the sample of mock galaxies to estimate uncertainties for the structural and photometric quantities measured in this work. In figures \ref{fig:kde_m},\ref{fig:kde_mu}, \ref{fig:kde_re} and \ref{fig:kde_n} we show the distributions of errors for each parameter $\rm p \in$ \{$\rm n,R_e,m,\langle \mu \rangle_e$\}, where \textit{error} means: $\rm \Delta p = p_{\rm true} - p_{\rm fitted}$. To estimate the uncertainties for a galaxy with fitted parameters $\rm n$, $ R_e$ and $\rm m$, we gather a sample of mock galaxies with similar fitted parameters, compute the median and median absolute deviation (MAD) for the errors of each parameter, and finally compute uncertainties as deviations from the absolute median error. The computation is as below:
\begin{equation}
    \rm \sigma_p = |\widetilde{X}_{\Delta p}| + MAD(X_{\Delta p}),
\end{equation}
where $\rm \sigma_p$ is the uncertainty of a parameter $\rm p$, $\rm {X}_{\Delta p}$ is a set of errors for that parameter and $\rm \widetilde{X}_{\Delta p}$ is the median of $\rm {X}_{\Delta p}$. We use MAD as deviation for the errors because we are not assuming that they are normally distributed. For each uncertainty estimated, we require at least 200 mock galaxies with similar fitted parameters, which means explicitly mock galaxies whose parameters are within: $\pm 0.25$ for $\rm n$, $\pm 25$ pc for $\rm R_e$ and $\pm 0.25$ mag for $\rm m$, from the fitted parameters of the real galaxy. For absolute magnitude and physical radius uncertainties, we also consider the uncertainty in the distance modulus of \cite{Tonry01}.

\section{Individual LSBds and LSBds GC systems}
\label{sec:indivual_plots}
\begin{figure*}
    \centering
    \includegraphics[width=0.5\linewidth]{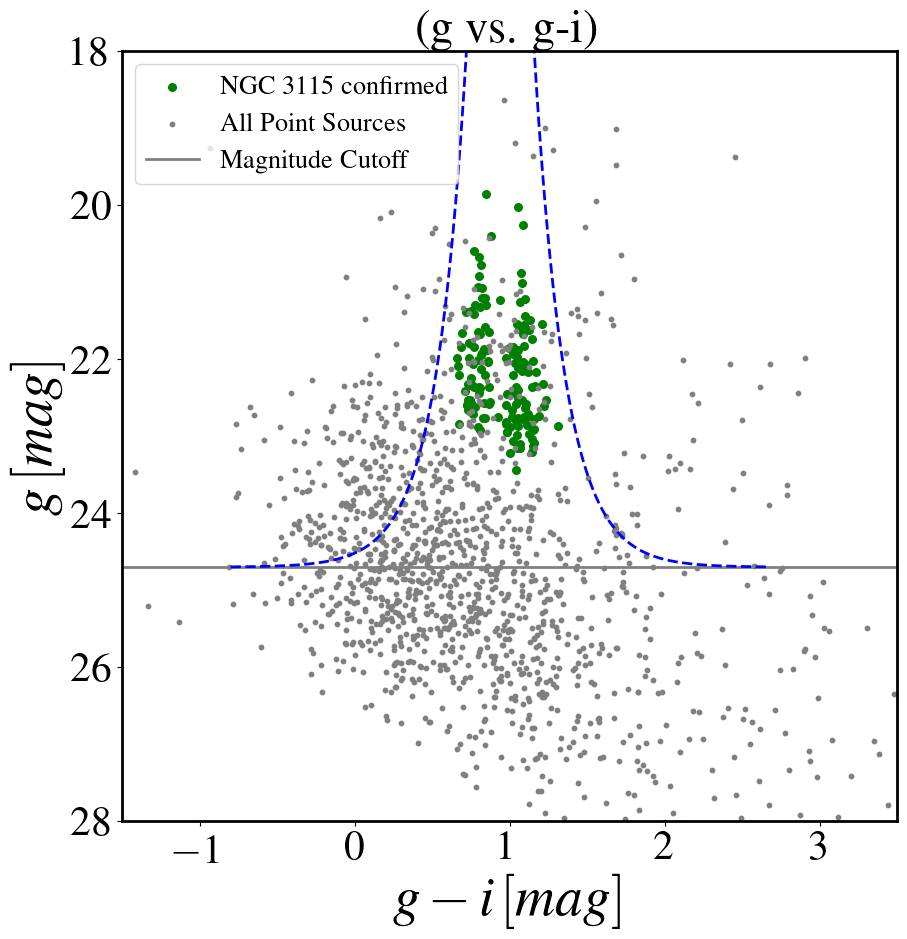}
    \caption{Colour magnitude diagram for GC candidates in all our sample galaxies. The dashed blue lines represent boundaries in point source selection in colour and the horizontal grey line the boundary selection in magnitude. The gray circles indicates all of the point sources detected in our sample. The green circles indicate the GCs associated with the NGC 3115 \citep{Forbes2017}.}
    \label{fig:bigCMDGCs}
\end{figure*}

In \autoref{fig:stamps} we present all of the individual image stamps, models and residuals of our sample LSBds. In a similar manner to \autoref{fig:13dist}, in \autoref{fig:gcs} we present the GC candidates selection for the complete sample of LSBds followed-up with Gemini/GMOS.
In \autoref{fig:bigCMDGCs} we show the candidate sources detected in the Gemini stamps in red. In blue, we show NGC\,3115 GCs from \cite{Forbes2017} and in grey GAIA IDR3 point sources found around the same region of each one of the LSB dwarf stamps (the same ones described in \autoref{sec:profile_fitting}).   
We select our candidate GCs using the following criteria: $MAG\_AUTO\_g \le 24.7$, $ 0 \le g-i  \le 1.75$, $0 \le g-z \le 2$ $ CLASS\_STAR\_g  \ge 0.05$ and keep the objects with galactocentric distances up to $3.5 R_{e}$.
These criteria are chosen considering the region where the GCs from NGC 3115 are found in previous works. \citep[e.g.][]{Faifer11,Forbes2017}.
Also we select objects with MAG\_AUTO\_g $\le 24.7$. This is $\sim2$ Magnitudes below the globular cluster luminosity function (GCLF). 
The vertical and horizontal lines in \autoref{fig:bigCMDGCs} represent our selection criteria.

\begin{figure*}
    \centering
    \includegraphics[width=0.9\linewidth]{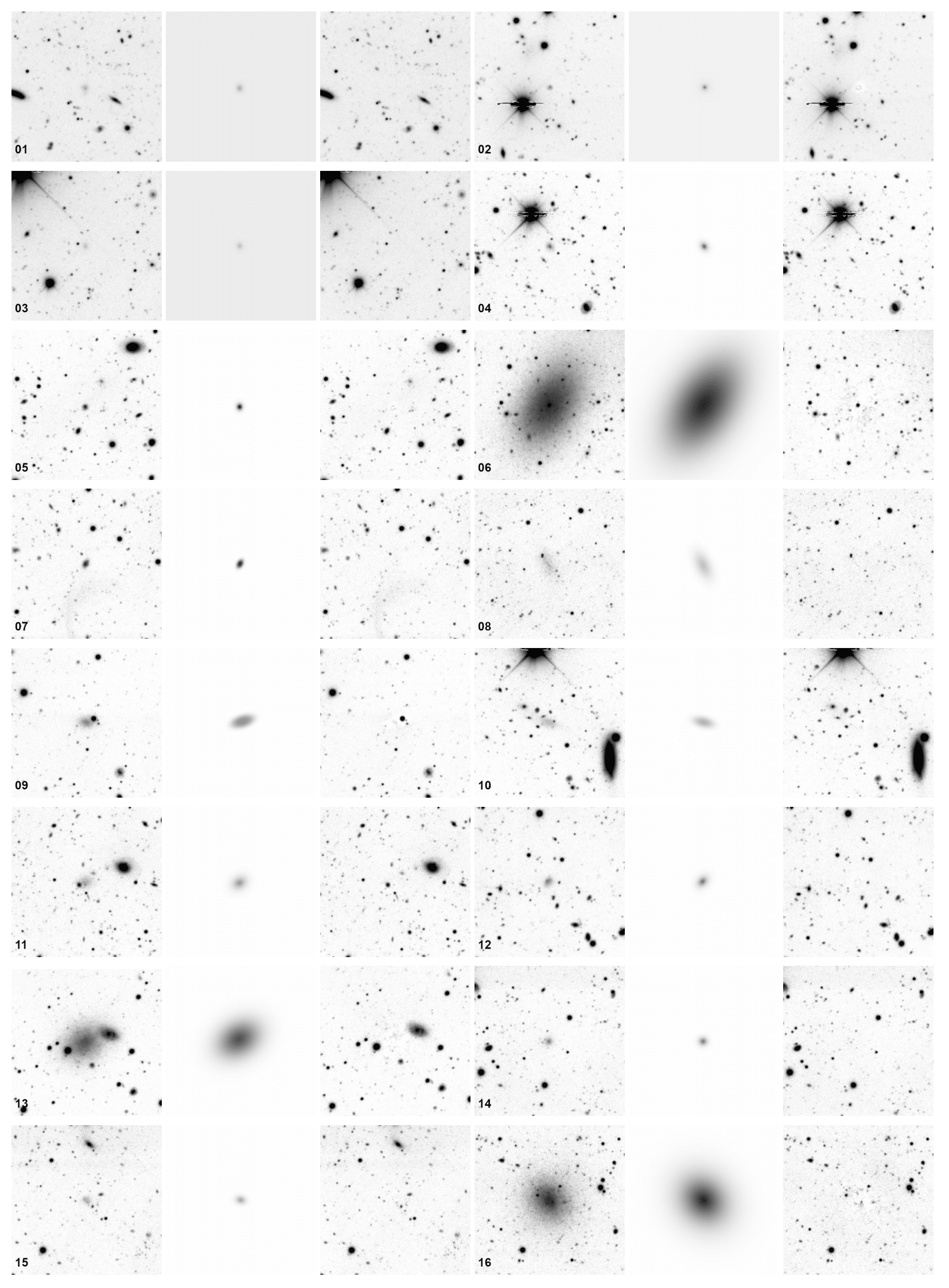}
    \caption{LSBs stamps according to the legend. From left to right stamp, model, residual. The size of the stamps are 400x400 ($\sim 108\arcsec \times 108\arcsec$). The image scale is ASINH, where on the low-limit of the scale parameter we used the sky level obtained with SExtractor}
  \label{fig:stamps}
\end{figure*}
\begin{figure*}
    \ContinuedFloat
    \centering
    \includegraphics[width=0.9\linewidth]{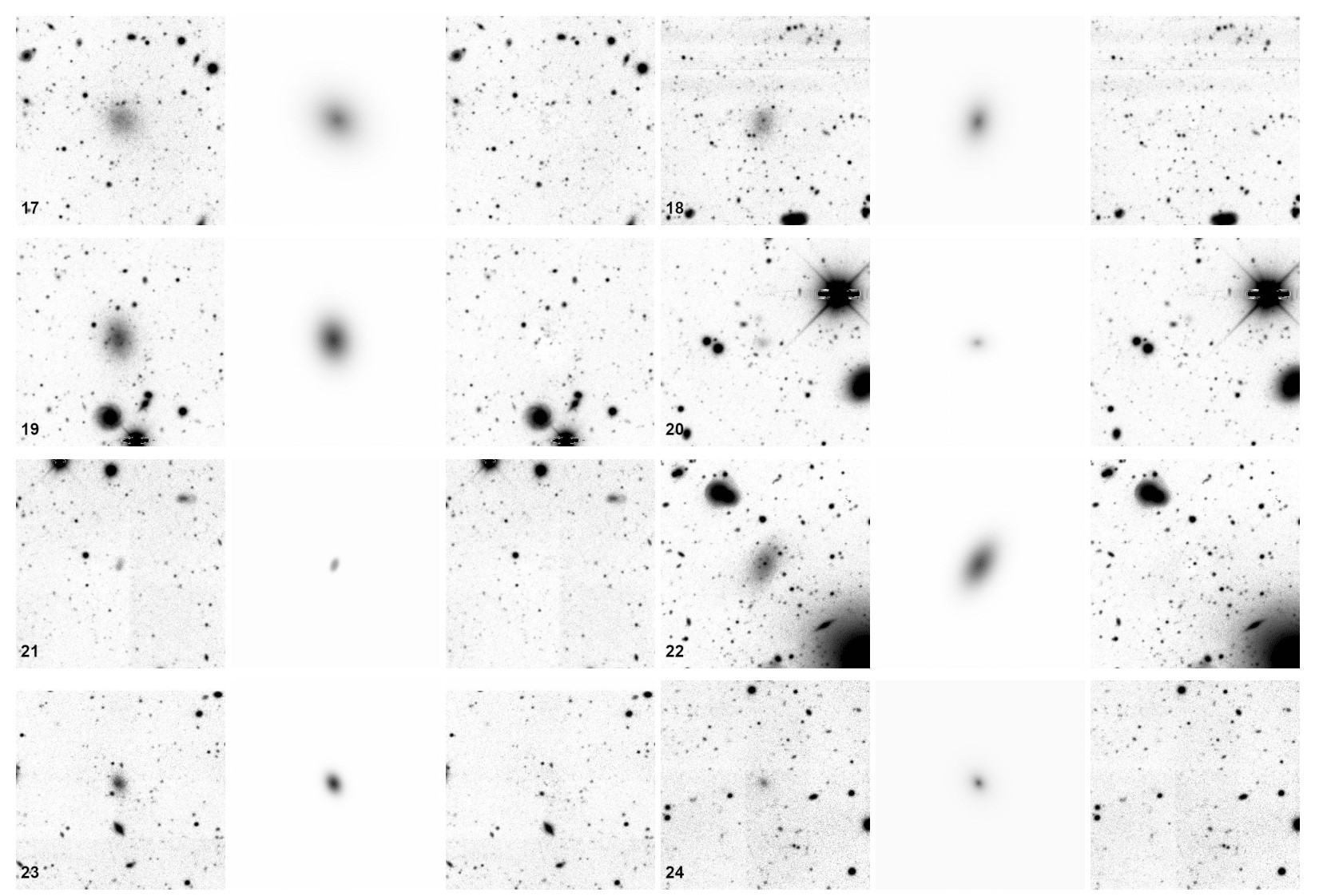}
    \caption{Continued.}
    %\label{fig:stamps}
\end{figure*}

\begin{figure*}
    \centering
    \includegraphics[scale=0.72]{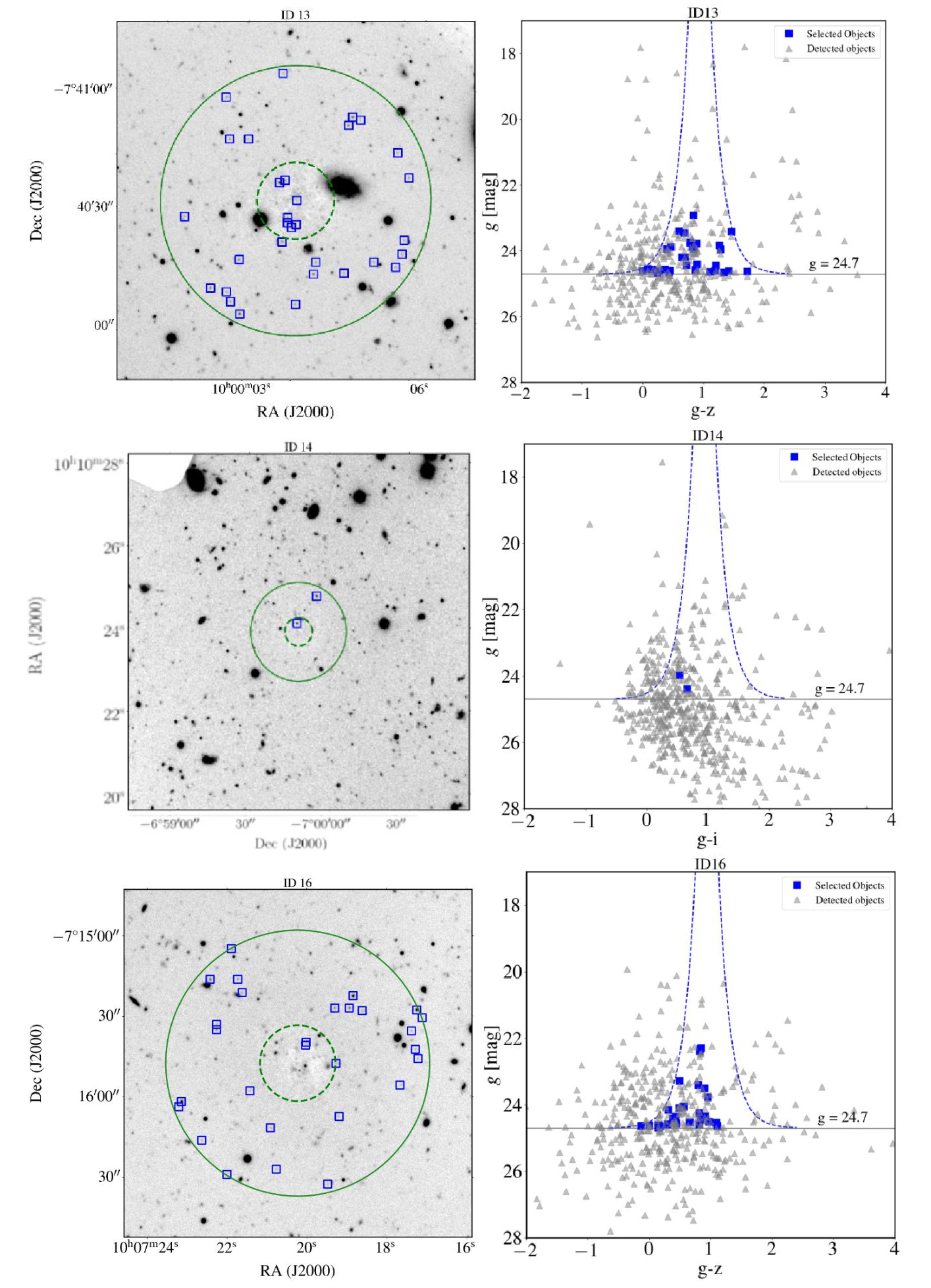}
     \caption{Same as Fig.\ref{fig:13dist} for all systems with GCs, according to the legend.}
     \label{fig:gcs}
\end{figure*}
\begin{figure*}
    \centering
         \includegraphics[scale=0.72]{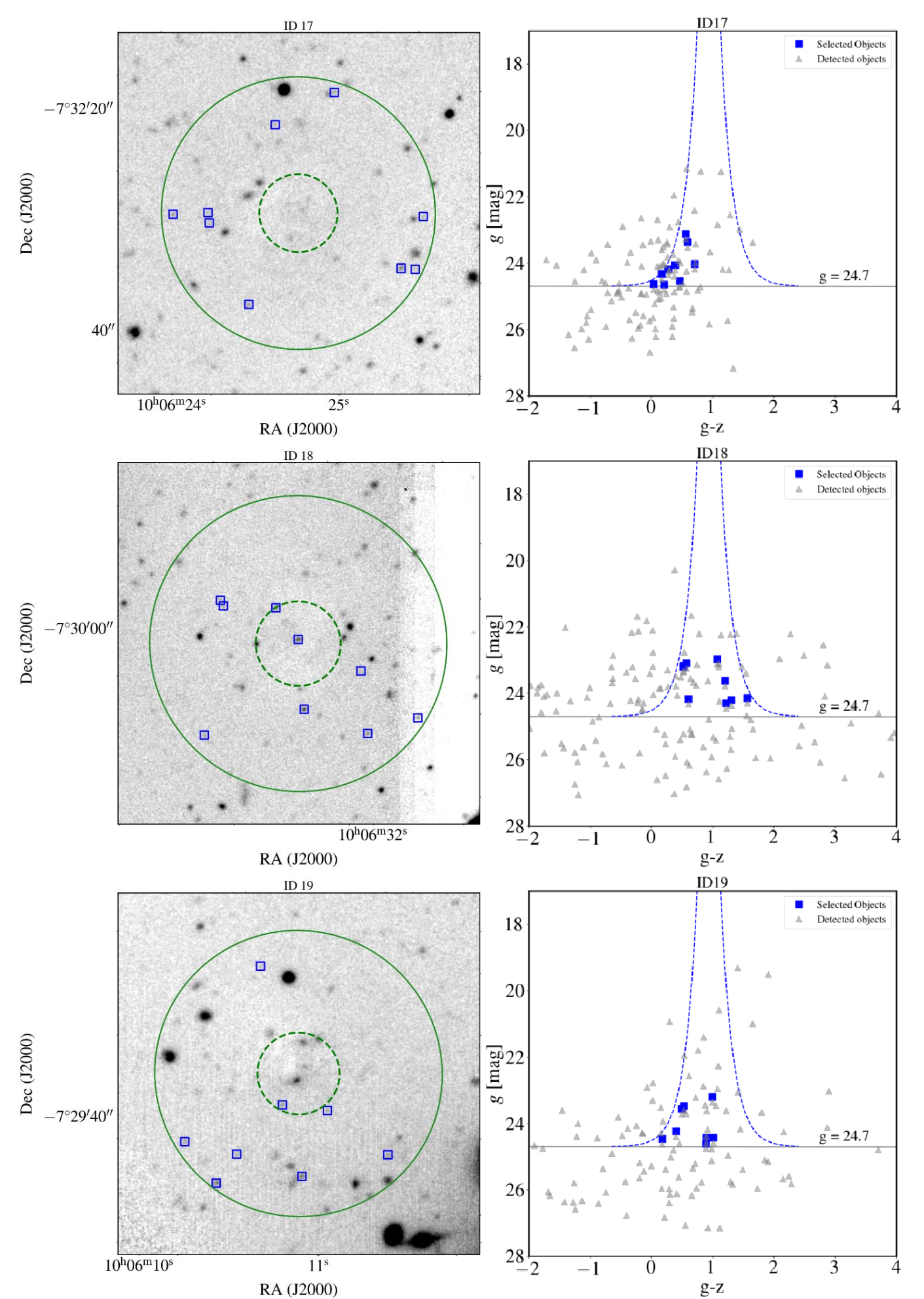}
     \caption{Continued.}
     \label{fig:gcs3}
 \end{figure*} 
 \begin{figure*}
    \centering
         \includegraphics[scale=0.7]{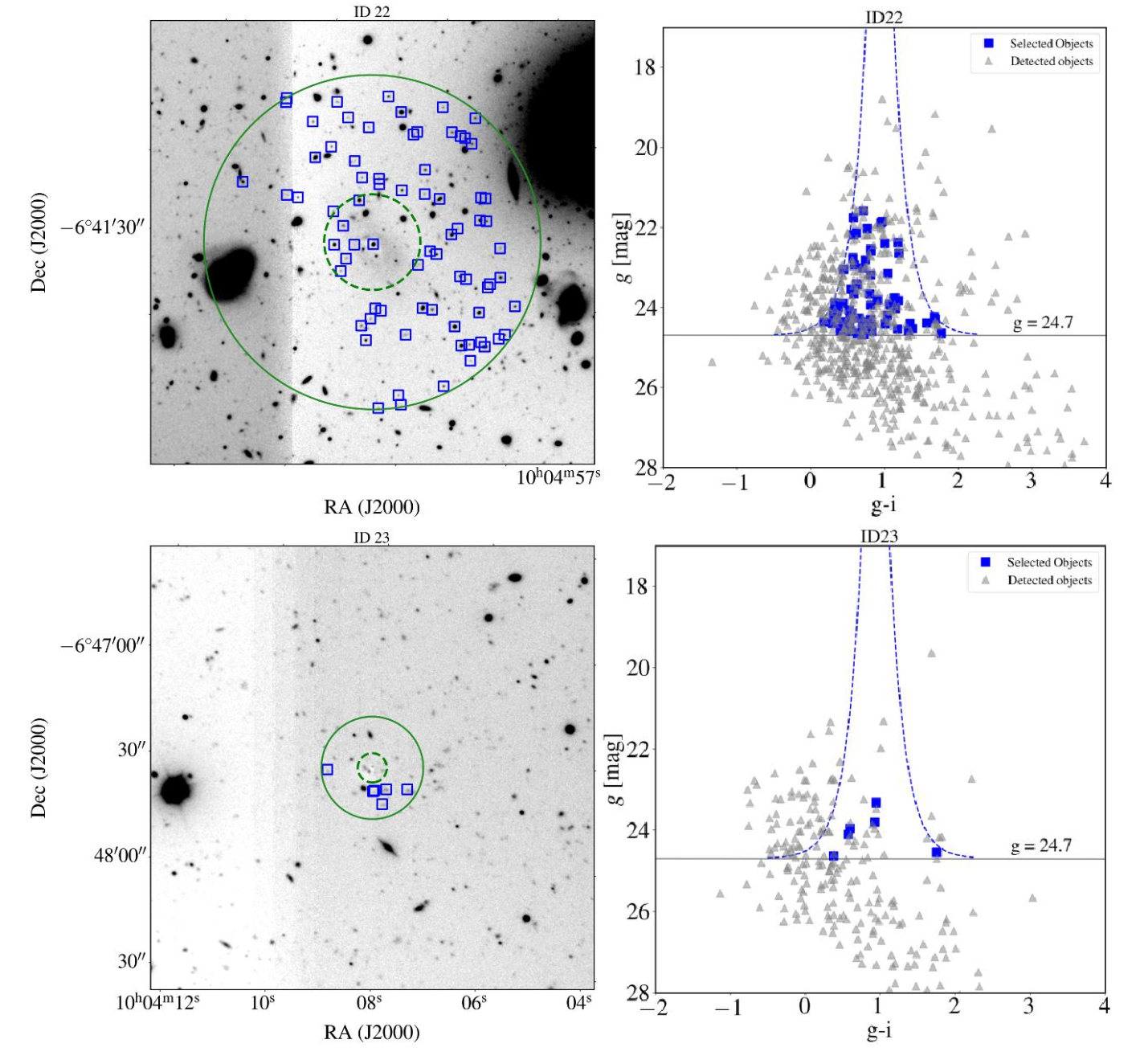}
     \caption{Continued.}
     \label{fig:gcs}
 \end{figure*}

% Don't change these lines
\bsp	% typesetting comment
\label{lastpage}
\end{document}